\documentclass[useAMS,usenatbib]{mnras}
\usepackage[pdftex]{graphicx}
\usepackage{xcolor}
\usepackage{hyperref}
\usepackage{amsmath}
\usepackage{amsfonts}
\usepackage{amssymb}
\usepackage{subcaption}
\usepackage{mwe}
\usepackage{enumitem}
\usepackage{multirow}

\newcommand{\kms}{km s$^{-1}$}
\def\q#1{`#1'}

\newcommand{\brick}{G0.253$+$0.016}
\newcommand{\ap}{$\sim$ }
\newcommand{\msun}{M$_{\odot}$}
\newcommand{\cmq}{cm$^{-3}$}
\newcommand{\asec}{$^{\prime}$$^{\prime}$}
\title[Star formation in \q{the Brick}]{Star formation in \q{the Brick}: ALMA reveals an active proto-cluster in the Galactic centre cloud \brick}
\author[D. L. Walker et al.]{Daniel~L.~Walker$^{1,2,3}$\thanks{E-mail: \texttt{daniel.walker.astro@gmail.com}},
Steven~N.~Longmore$^{4}$,
John Bally$^{5}$,
\and Adam Ginsburg$^{6}$,
J.~M.~Diederik~Kruijssen$^{7}$,
Qizhou Zhang$^{8}$,
\and Jonathan~D.~Henshaw$^{9}$,
Xing Lu$^{2}$, 
João Alves$^{10,11}$,
Ashley~T.~Barnes$^{12}$,
\and Cara Battersby$^{1}$, 
Henrik Beuther$^{9}$,
Yanett~A.~Contreras$^{13}$,
Laura G\'{o}mez$^{3}$,
\and Luis~C.~Ho$^{14,15}$,
James~M.~Jackson$^{16}$,
Jens Kauffmann$^{17}$,
Elisabeth~A.~C.~Mills$^{18}$,
\and and Thushara Pillai$^{19}$\vspace{0.2cm}\\
$^{*}$Author affiliations are listed at the end of the paper}

\begin{document}
\date{}
\pagerange{\pageref{firstpage}--\pageref{lastpage}} \pubyear{2021}
\maketitle
\label{firstpage}
\begin{abstract}
\brick, aka \q{the Brick}, is one of the most massive (\textgreater \ 10$^{5}$~\msun) and dense (\textgreater \ 10$^{4}$~\cmq) molecular clouds in the Milky Way's Central Molecular Zone. Previous observations have detected tentative signs of active star formation, most notably a water maser that is associated with a dust continuum source. We present ALMA Band 6 observations with an angular resolution of 0.13\asec \ (1000~AU) towards this \q{maser core}, and report unambiguous evidence of active star formation within \brick. We detect a population of eighteen continuum sources (median mass \ap 2~\msun), nine of which are driving bi-polar molecular outflows as seen via SiO (5-4) emission. At the location of the water maser, we find evidence for a protostellar binary/multiple with multi-directional outflow emission. Despite the high density of \brick, we find no evidence for high-mass protostars in our ALMA field. The observed sources are instead consistent with a cluster of low-to-intermediate-mass protostars. However, the measured outflow properties are consistent with those expected for intermediate-to-high-mass star formation. We conclude that the sources are young and rapidly accreting, and may potentially form intermediate and high-mass stars in the future. The masses and projected spatial distribution of the cores are generally consistent with thermal fragmentation, suggesting that the large-scale turbulence and strong magnetic field in the cloud do not dominate on these scales, and that star formation on the scale of individual protostars is similar to that in Galactic disc environments.
\end{abstract}

\begin{keywords}
Stars: formation -- ISM: clouds -- Galaxy: centre
\end{keywords}

\section{Introduction}\
\label{sec:intro}
The Milky Way's Central Molecular Zone (CMZ, inner few hundred parsecs) contains a substantial reservoir (\textgreater \ 10$^{7}$~\msun) of dense (\textgreater \ 10$^{4}$~\cmq) molecular gas \citep{CMZ}. Despite this, the star formation rate (SFR) in the CMZ is at least an order of magnitude lower than predicted by star formation relations that have been calibrated in nearby galactic disc environments \citep{snl_sf}. This relative dearth of star formation is observed both on global scales and on the scales of individual molecular clouds in the CMZ \citep[e.g.][]{Barnes_17, Kauffmann17b, Lu19}. This deviation from the expected star formation rate is important, as it suggests that the criteria required for stars to form varies as a function of environment. If this is true, then it is crucial that this variation is understood and characterised, such that star formation relations can be accurately applied to the varying environmental conditions found throughout the Universe. 

\begin{figure}
\begin{center}
\label{fig:brick_3col}
\includegraphics[scale=0.38, angle=0]{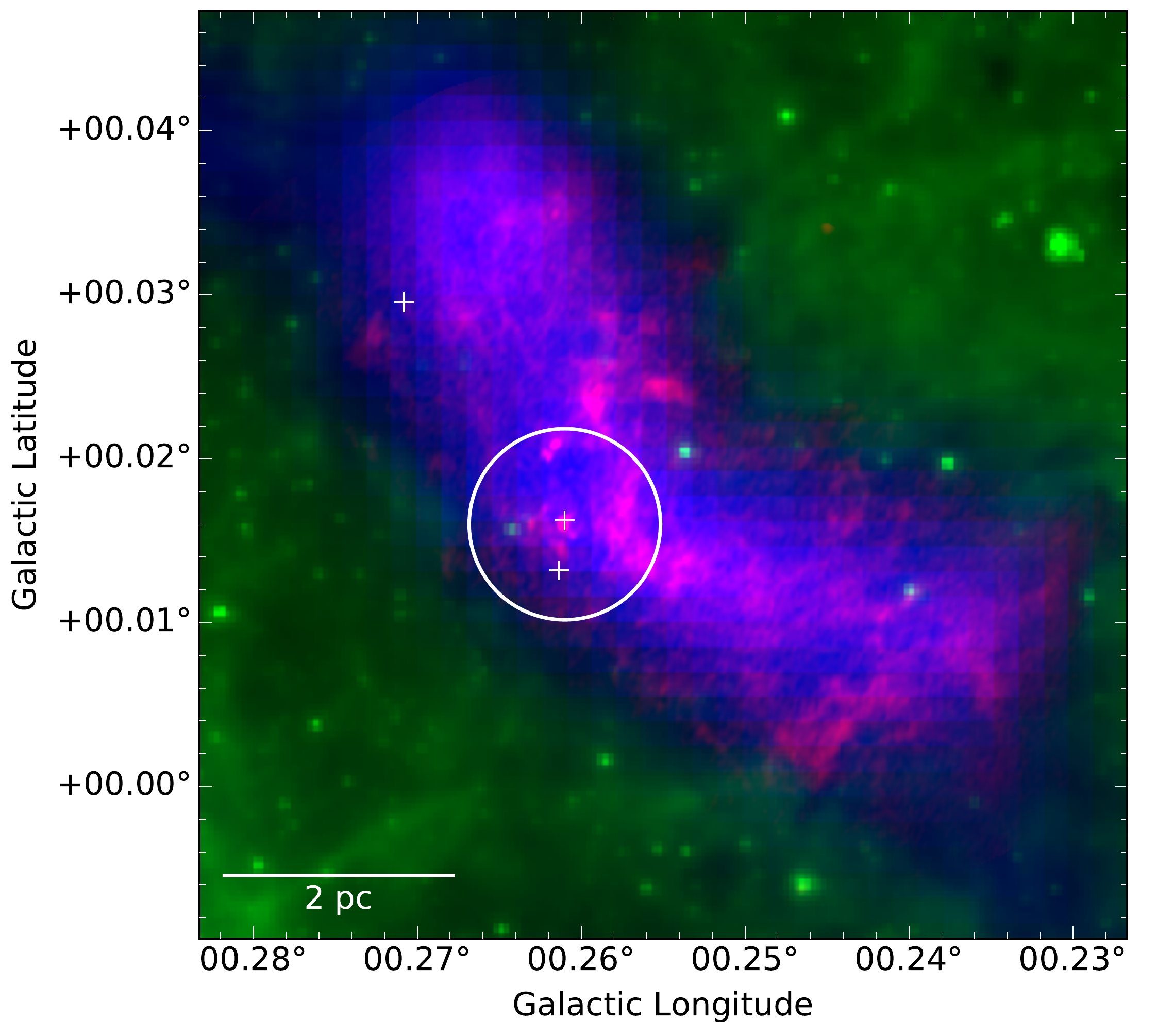}
 \caption{Three-colour image of \brick. {\bf{Red}}: ALMA 3~mm dust continuum \citep{Jill_pdf_2014}, {\bf{Green}}: Spitzer/GLIMPSE 8~$\mu$m emission \citep{glimpse}, {\bf{Blue}}: Herschel/HiGAL dust column density \citep{Cara_higal, higal_new}. The white crosses indicate the positions of known water masers \citep{Lu19}. The white circle corresponds to the primary beam field of view of the ALMA observation reported in this paper.}
\end{center}
\end{figure}

While the CMZ appears to be under-producing stars as a whole relative to the amount of dense gas it contains, one molecular cloud in particular has been the focus of significant research efforts in this context. \brick \ (also known as \q{the Brick}) stands out as an extreme infra-red dark cloud against the intense mid-IR background (see Figure \hyperref[fig:brick_3col]{1}). The cloud contains \textgreater \ 10$^{5}$~\msun \ of material within a mean radius of only a few parsecs \citep[2-3~pc, e.g.][]{Immer, Brick_snl, Walker15}. Yet despite this substantial reservoir of dense material, no evidence of embedded star formation has been observed in the cloud other than a water maser that coincides with a compact millimetre continuum source \citep[e.g.][]{Lis_brick1, Immer, Brick_jens, Brick_Katharine, Jill_pdf_2014, Mills15, Lu19}. Hereafter, we refer to this source as the \q{maser core} for brevity. We note that there are at least two more water masers in \brick \ (see Figure \hyperref[fig:brick_3col]{1}), however, no counterparts have been detected in the dust continuum \citep{Lu19}.

Deep radio continuum observations and further searches for maser emission do not reveal any additional signatures of embedded star formation towards this source \citep[e.g.][]{Immer, Rod13, Mills15, Lu19B}. Potentially embedded star formation has been inferred in \brick \ due to the presence of warm dust along one edge of the cloud \citep{Marsh16}. \citet{Lis2001} also suggest that the internal luminosity of the cloud could correspond to the presence of \ap four B0 zero-age main-sequence stars. Another potential indication of star formation is the detection of an arc-like structure in the cloud that is close to the maser core in projected position \citep{Higuchi14, Mills15, Henshaw_2019}. Though the origin of this structure has been disputed, new results suggest that it could be a feedback-driven shell of material, which may indicate embedded star formation (Henshaw et al. in prep.).

These properties make \brick \ one of the most massive and dense molecular clouds known to exist in the Galaxy in which there are no unambiguous signs of widespread star formation. The lack of on-going star formation in \brick, coupled with similar evidence in other CMZ clouds, has been argued to favour an environmentally-dependent critical density threshold for star formation \citep[e.g.][]{Jill_pdf_2014, Diederik_cmz_sf, Walker18, Ginsburg18, Barnes_2019}. It has been proposed that the CMZ undergoes an episodic cycle of star formation, and is currently at a low point due to the high turbulent energy there \citep{Diederik_cmz_sf, krumholz15, Krumholz17, Armillotta19}.

The high turbulent energy is evidenced observationally as broad line-widths of \ap 10 -- 20~km s$^{-1}$ on large (parsec) scales \citep{Henshaw_cmz}. This high turbulence will act to drive up the critical volume density threshold for star formation \citep[e.g.][]{krumholz05, Padoan11, Federrath12, Hennebelle13, Padoan14}, and may therefore explain the discrepancy between the observed current SFR and predictions based upon proposed density thresholds \citep[e.g.][]{Lada10, Lada_2012_sfr}. Recent results from a high-resolution survey of the CMZ using the Submillimeter Array, \emph{CMZoom}, show that there is an overall lack of compact substructure within the the dense CMZ clouds, which is likely due to their inability to form such structure in this turbulent environment \citep{cmzoom1, cmzoom2}. 

\citet{Federrath16} explored this in \brick \ specifically, and concluded that the turbulence in the cloud is likely dominated by solenoidal turbulence, which is driven by the strong shear in the CMZ's deep gravitational potential \citep{kruijssen19} and could suppress the SFR by a factor of several \citep{dale19}. The strong (\ap mG) magnetic field in \brick \ has also been discussed as a potential source of support, which could suppress fragmentation and thus star formation in the cloud \citep{Pillai_brick_bfield}. Given their relative proximity \citep[\ap 8.1~kpc,][]{Gravity19, Reid_2019}, clouds such as \brick \ therefore offer ideal laboratories in which we can study the process of star formation in an extreme, turbulent environment, on scales that are otherwise inaccessible in extragalactic analogues.

\brick \ also presents an ideal region in which to search for the precursors to high-mass stars (\textgreater \ 8~\msun) and massive stellar clusters (\textgreater \ 10$^{3}$~\msun). The fact that the cloud contains \textgreater \ 10$^{5}$ \msun \ within a mean radius of \ap 3~pc, yet  has no apparent widespread star formation, has led to the proposal that we could be witnessing the initial conditions of massive cluster formation \citep{Brick_snl, Brick_jill, rathborne15, Walker15, Walker16}, though the star forming potential of the cloud has been debated \citep{Brick_jens}. Assuming a star formation efficiency (SFE) of 10 -- 30\%, \brick \ has the potential to form a \ap 10$^{4}$~\msun \ cluster. If the cloud were to ultimately form such a massive cluster, then a statistical argument would also suggest the likely presence of precursors to high-mass stars due to significant sampling of the stellar initial mass function (IMF). Indeed, it is known that the CMZ harbours several young massive stellar clusters, such as the Arches and Quintuplet, that contain many high-mass stars, and even some extremely massive stars \citep[\textgreater \ 100~\msun, e.g.][]{Figer_quintuplet, Figer_arches2}. Given that \brick \ is one of the best candidates for representing a quiescent precursor to such clusters, it therefore follows that it is a good candidate for hosting the initial conditions for massive star formation.

While the \q{maser core} in \brick \ constitutes the best evidence for potentially active star formation within the cloud, the source has not been found to coincide with any 70~$\mu$m point sources, radio continuum emission, nor any significant molecular line emission that would indicate the presence of hot cores \citep[e.g.][]{Brick_jens, Jill_pdf_2014}. In this paper, we present high angular resolution Atacama Large Millimeter/submillimeter Array (ALMA) observations of this \q{maser core} in \brick. These observations reveal the presence of fragmentation, bipolar outflows and internal heating -- unambiguous confirmation of active star formation in \brick. Section \hyperref[sec:observations]{2} gives an overview of the observations and imaging techniques used. Section \hyperref[sec:results]{3} presents the results of the observations: (\hyperref[sec:continuum]{i}) the 1.3~mm dust continuum and the physical properties of the detected sources, and (\hyperref[sec:lines]{ii}) the molecular line emission, specifically from SiO (5-4), $^{13}$CO (2-1), and CH$_{3}$CN J=12-11. Section \hyperref[sec:discussion]{4} provides a discussion of the results and the implications for our understanding of star formation both in \brick \ and the CMZ in general.

\section{Observations \& Data Reduction}\
\label{sec:observations}
\begin{table*}
\begin{center}
\label{tab:observations}
\caption{Details of the three observed execution blocks. Listed are the observation dates, nominal array configurations, number of 12~m antennas in the array, full range of antenna baseline lengths, atmospheric precipitable water vapour content (PWV), total time on source, and the bandpass, flux, and phase calibrators used for each observation.}
  \begin{tabular}{cccccccccc}
    \hline
    Date & Array & Antennas & Baselines & PWV & Time on source & Bandpass & Flux & Phase \\
    (d/m/y) & configuration & \# & (m) & (mm) & (minutes) & calibrator & calibrator & calibrator \\ \hline
    25/04/2017 & C40-3 & 40 & 15 -- 459 & 1.00 & 27.82 & J1924-2914 & Titan & J1744-3116\\
    19/07/2017 & C40-6 & 40 & 18 -- 3696 & 0.49 & 49.45 & J1924-2914 & J1733-1304 & J1744-3116\\
    21/07/2017 & C40-6 & 40 & 16 -- 3696 & 0.81 & 46.42 & J1924-2914 & J1733-1304 & J1744-3116\\
    \hline 
    \hline
  \end{tabular}
\end{center}
\end{table*}

\begin{table}
\begin{center}
\label{tab:spws}
\caption{Overview of the spectral setup used for our ALMA observation. The specific line(s) targeted per spectral window are given, along with the corresponding central frequency ($\nu_{\textrm{cent}}$), bandwidth (BW), and spectral resolution in terms of velocity ($\Delta v$). While these are the lines that were specifically chosen, there are many more lines observed within these spectral windows.}
  \begin{tabular}{ccccc}
    \hline
    Spectral & $\nu_{\textrm{cent}}$ & BW & $\Delta v$\\
    window & (GHz)	& (GHz) & (km~s$^{-1}$)\\ \hline
    SiO (5-4) & 217.105 & 0.234 & 0.78\\
    H$_{2}$CO (3$_{0,3}$ -- 2$_{0,2}$) & 218.222 & 0.234 & 0.78\\
    H$_{2}$CO (3$_{2,2}$ -- 2$_{2,1}$) & 218.476 & 0.234 & 0.78\\
    H$_{2}$CO (3$_{2,1}$ -- 2$_{2,0}$) & 218.760 & 0.234 & 0.78\\
    $^{13}$CO (2-1)/CH$_{3}$CN (12-11) & 220.709 & 0.934 & 0.77\\
    Continuum & 232.500 & 1.875 & 2.50\\
    Continuum & 235.000 & 1.875 & 2.47\\    \hline 
    \hline
  \end{tabular}
\end{center}
\end{table}

We obtained high-sensitivity, high-angular-resolution dust continuum and molecular line observations towards the \q{maser core} in \brick \ with ALMA at \ap 230~GHz (Band 6, 1.3~mm) as part of the Cycle 4 project 2016.1.00949.S (PI: D. Walker). The observations were taken as a single pointing centred on the source (G0.261$+$0.016, see Figure \hyperref[fig:brick_3col]{1}), using only the main 12~m array. The correlator was configured to target 7 spectral windows, 5 of which targeted specific molecular transitions in the lower sideband with a spectral resolution of \ap 0.77 \kms. The remaining 2 spectral windows were dedicated to broad-band continuum detection in the upper sideband, with a spectral resolution of \ap 2.5 \kms. The total aggregate bandwidth is approximately 5.6~GHz. The project was observed across 3 individual execution blocks between April and July 2017. Each execution used 40 antennas, with baselines ranging from 15 -- 3696~m. Full details concerning the observations and spectral setup are given in Tables \hyperref[tab:observations]{1} \& \hyperref[tab:spws]{2}, respectively. 

\subsection{Imaging}
\label{sec:imaging}
The ALMA pipeline calibrated data sets for each execution block were combined to obtain final data products, which were then imaged in CASA \citep{casa}. Prior to final imaging, dirty cubes were created for each spectral window, and ran through the \texttt{findContinuum}\footnote{\url{https://almascience.nrao.edu/documents-and-tools/alma-science-pipeline-users-guide-casa-5-6.1}} routine in CASA in pipeline mode to determine the continuum-only channels in each window \citep{Humphreys16}. The continuum was then imaged in \texttt{tclean} by combining all spectral windows and specifying the previously identified channels to be considered when generating the continuum. The final continuum image that is used throughout this paper was imaged using the Briggs weighting scheme with a robust parameter of 0.5, multi-scale deconvolution, and with the \texttt{auto-multithresh}\footnote{\url{https://casaguides.nrao.edu/index.php/Automasking\_Guide}} masking option \citep{Automultithresh}, using the default auto-masking parameters. The resultant image has a synthesised beam size of 0.17\asec \ $\times$ 0.12\asec \ (\ap 1400~AU$ \times$ 1000~AU), with a continuum sensitivity of \ap 50~$\mu$Jy~beam$^{-1}$ (0.07~K). 
The largest angular scale is \ap 10\asec \ (0.4~pc).

Along with the continuum, we also imaged all spectral windows to produce full data cubes. The cubes were imaged using mostly the same parameters as for the continuum, with the exception of a higher cleaning threshold, and an auto-masking \texttt{negativethreshold} parameter of 7.0 (default is 0.0) to account for any absorption. In contrast to the continuum, we opted to perform the cleaning prior to continuum subtraction. This was done as we found that using the \texttt{uvcontsub} task prior to cleaning did not perform a satisfactory continuum subtraction for the more line-rich spectral windows. Instead, we used the {\sc statcont}\footnote{\url{https://hera.ph1.uni-koeln.de/~sanchez/statcont}} Python package, which is specifically designed to determine the continuum level in line-rich data and perform continuum subtraction \citep{statcont}. The resulting line sensitivity in a 0.78~\kms \ channel is \ap 850~$\mu$Jy~beam$^{-1}$ (1.25~K).

\section{Results}
\label{sec:results}
\subsection{Continuum Data}\
\label{sec:continuum}

\begin{figure*}
\begin{center}
\label{fig:full_continuum}
\includegraphics[scale=0.63, angle=90]{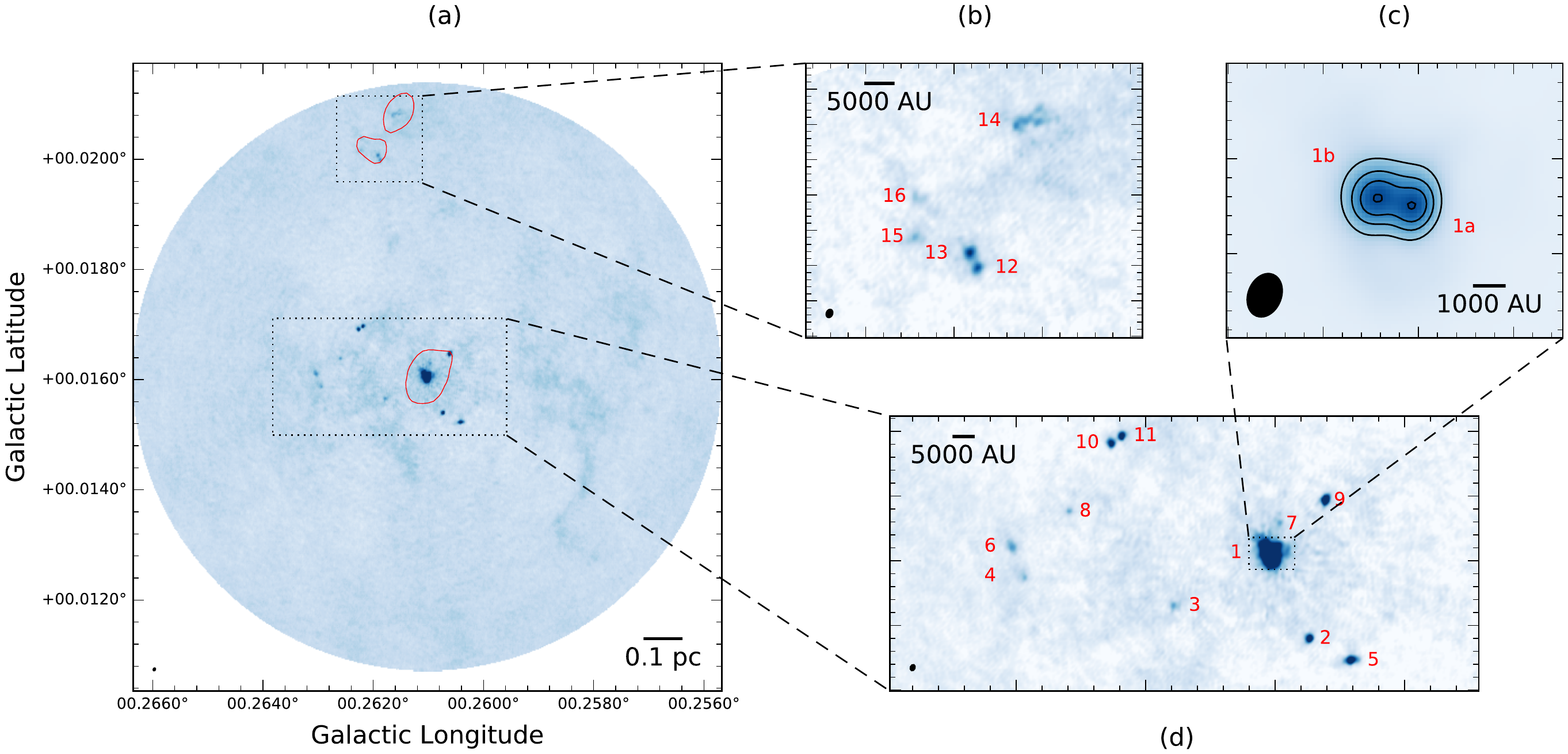}
\vspace{-65pt}
 \caption{The 1.3~mm dust continuum image towards the \q{maser core} in \brick \ as seen with ALMA at an angular resolution of 0.13\asec (\ap 1000~AU). {\bf{2(a)}} displays the full field. The red contours show the 3mm dust continuum from \citep{Jill_pdf_2014}, and the dotted boxes highlight the zoom-in regions shown in sub-figures (b) and (d). {\bf{2(b)}} and {\bf{2(d)}} show zoomed-in images of the compact sources detected via dendrograms. {\bf{2(c)}} shows a zoom-in of the bright central source, denoted as core \q{1}, overlaid with contours of [3, 5, 7, 9]~mJy beam$^{-1}$ or [4.4, 7.4, 10.3, 13.3]~K.}
\end{center}
\end{figure*}

Figure \hyperref[fig:full_continuum]{2a} displays the 230~GHz continuum image of the full ALMA field. This observation reveals that, while the target field is still dominated by a bright central source on \ap 1000~AU scales, there is a clear population of fainter compact sources. To quantify this substructure, we compute dendrograms using the {\sc astrodendro} Python package. In brief, dendrograms are hierarchical clustering algorithms, in which structure in a data set is represented as a \q{tree}, where substructures are classified as \q{branches}, and local maxima at the highest level of the branch structures are called \q{leaves}. Using this nomenclature in the context of our continuum data, each \q{leaf} represents a continuum source or core.

To compute the dendrogram, a threshold of 3$\sigma$, an increment between structures of 1$\sigma$, and a minimum number of pixels in a source of 100 are specified (which is \ap 50\% of the synthesised beam), where $\sigma$ \ap 50~$\mu$Jy~beam$^{-1}$. The number of sources and their properties are not strongly dependent on the choice of parameters, with the exception of the central source, which is more extended. We discuss the nature of the central source and its embedded structure later in this section. A total of 17 compact continuum sources are detected using dendrograms, which are highlighted in the zoom-ins in Figure \hyperref[fig:full_continuum]{2(b - d)}. The general properties of these sources are presented in Table \hyperref[tab:cores]{3}, including their integrated fluxes, sizes and estimated masses.

Assuming that the 1.3~mm continuum flux arises from optically thin dust emission (which is likely justified, see \citealt{Lu19B}), the masses of the detected sources are estimated using the following equation:
\begin{equation}
\label{eq:mass}
M = \frac{d^2}{\kappa_{\nu} B_{\nu}(T)} \int I_{\nu} d\Omega = \frac{d^2 F_{\nu}}{\kappa_{\nu} B_{\nu}(T)}
\end{equation}

\noindent where $M$ is the mass, $B_{\nu}$ is the Planck function, $T$ is the dust temperature, $\kappa_{\nu}$ is the dust opacity, $F_{\nu}$ is the integrated flux and $d$ is the distance. The dust opacity ($\kappa_{\nu}$) is not observationally constrained here, and so we estimate this using $\kappa_{\nu}=\kappa_{0}(\nu/\nu_{0})^{\beta}$, where $\kappa_{0}$ is taken to be 0.9~cm$^{2}$~g$^{-1}$ at $\nu_{0}=$ 230~GHz \citep{opacities}, and $\beta$ is assumed to be 1.75 \citep{Cara_higal}. The distance is assumed to be 8.1~kpc \citep{Gravity19, Reid_2019}. We note that \citet{Zoccali21} recently reported a distance of 7.2~kpc based on near-infrared star counts towards the cloud. If true, this would have the effect of decreasing our mass estimates by \ap 20\%. However, there are relevant caveats, particularly concerning the complicated interstellar extinction towards the Galactic centre, that must be investigated to further assess the validity of this result. Thus, we use the commonly assumed distance of 8.1~kpc for all analyses presented in this paper. 

The above contains the common assumption that the gas-to-dust ratio is 100, though this may not necessarily hold true in the CMZ \citep[e.g.][]{snl_sf, giannetti17}. The only remaining unknown in Equation \hyperref[eq:mass]{1} is the dust temperature. The dust temperature on these spatial scales is observationally unconstrained towards this source. Given the relatively large distance to the CMZ, measurements of the dust temperature are on \ap 30\asec \ scales from Herschel (Battersby et al. in prep). The average dust temperature towards this source is 22~K, and this is the value used in the estimation of the dust masses. We acknowledge that the masses reported here contain these uncertainties, and we explore the possibility of constraining these masses further with gas temperature estimates in section \hyperref[sec:gas_temp]{3.5}. However, we find no evidence for significant line emission towards the majority of sources in our field, which suggests that they are likely not significantly heated internally. This does not mean that the assumed dust temperature of 22~K is correct, but rather it is the best, and only measurement that we have for the majority of the sources.

Taking the aforementioned assumptions, we find that the sources range in mass from \ap 0.6 -- 64~\msun, with a median of 2~\msun. In addition to our assumptions, these masses are potentially lower limits due to the fact that the large scale emission, some of which may be associated with the cores, is filtered out by the interferometer. 

We also note that new results from the AzTEC survey of the CMZ measure higher values of $\beta$ of \ap 2 -- 2.5 towards \brick \ on scales of 10.5\asec \ \citep{aztec1, aztec2}. Substituting the upper value of this range over our assumed value of 1.75 would increase our reported masses by a factor of \ap 1.02. \citet{Marsh17} also use Herschel data to create higher resolution maps (12\asec) using the PPMAP procedure. The average dust temperature using this technique reduces to \ap 17~K. Assuming this value combined with $\beta$ =  2.5 would increase our reported dust continuum masses and densities by a factor of 1.42.

Although dendrograms pick out the central source as a large (R \ap 5000~AU) single object, manual inspection of the continuum data reveals further substructure. Figure \hyperref[fig:full_continuum]{2c} shows a zoom-in of the continuum emission from the central source overlaid with continuum contours. This reveals that the source is actually double-peaked, and suggests that it may be a protostellar binary/multiple system, with a projected separation between the two peaks of \ap 1000~AU. As this separation is approximately equivalent to the size of the synthesised beam, the sources are not well resolved. The peak intensities of the two sources are identical, at 9.1~mJy~beam$^{-1}$, suggesting that they may be of similar mass, assuming equal temperatures. A 2D Gaussian fit to the central objects yields deconvolved mean radii of \ap 1300~AU for both sources, and integrated fluxes of 12.1~mJy and 14.3~mJy, respectively for \#1a and \#1b (16.2~\msun and 18.0~\msun \ at 22~K). This would suggest that this is a massive protostellar binary, but there is a large uncertainty in the mass estimates due to the lack of dust temperature measurements. In the following section, we demonstrate that these two sources are internally heated, and are therefore likely less massive than the aforementioned estimates.

\begin{figure*}
\begin{center}
\label{fig:robust_minus2}
\includegraphics[scale=0.45, angle=0]{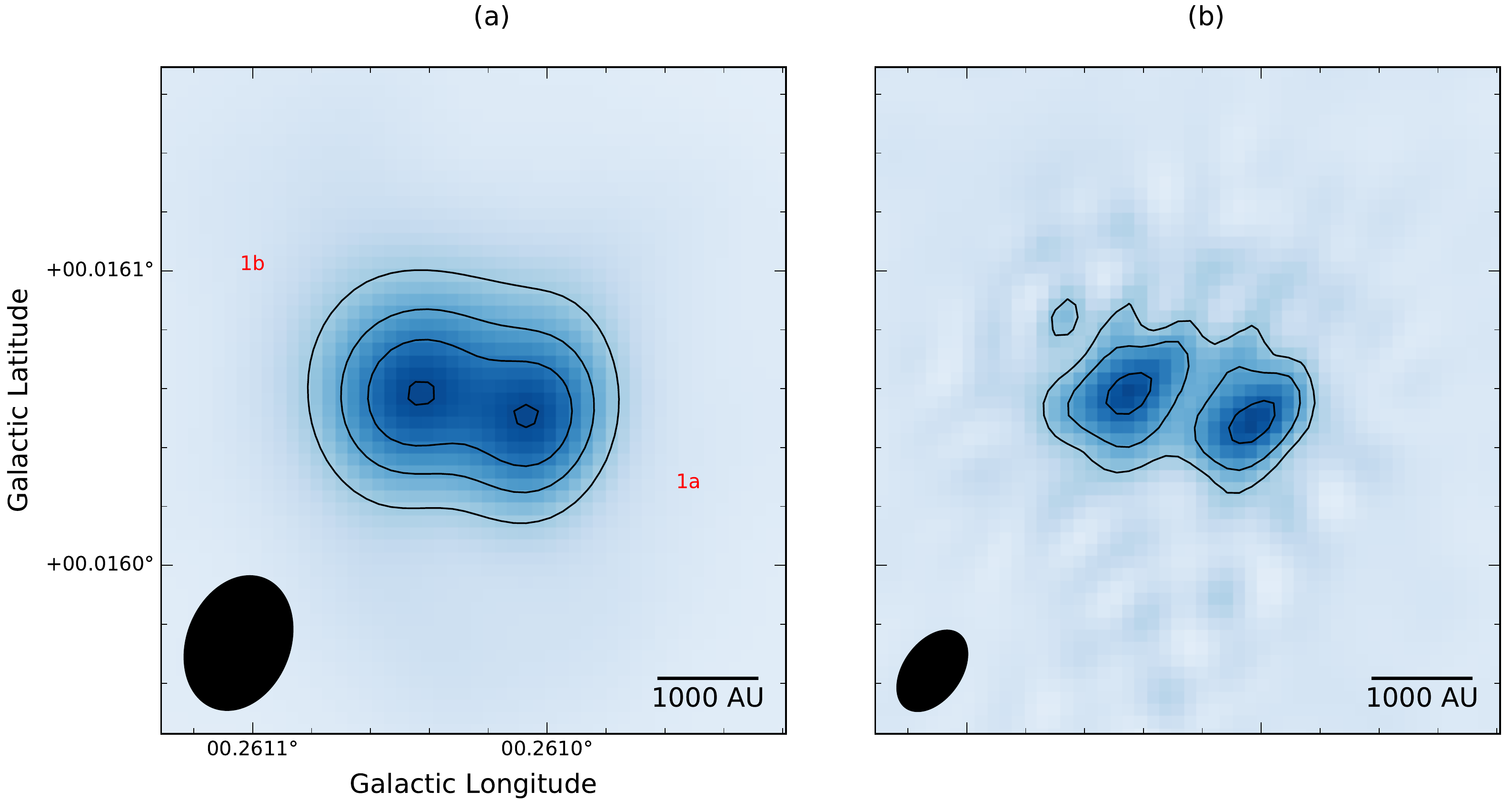}
 \caption{Comparison of the central region of our ALMA field showing the 1.3~mm dust continuum generated using the cleaning parameter \texttt{robust = 0.5} \textbf{(a)} and \texttt{robust = -2.0} \textbf{(b)}. Contours are the same as those in Figure \hyperref[fig:full_continuum]{2c}.}
\end{center}
\end{figure*}

To more clearly resolve this region, we also imaged the continuum using the Briggs weighting scheme with a robust parameter of -2.0 (i.e. uniform weighting), which prioritises resolution over sensitivity. The resulting image is shown in Figure \hyperref[fig:robust_minus2]{3}. Note that this image is not used for any analyses -- all results reported use the image generated with a robust parameter of 0.5. Using this weighting scheme we see that the central sources are more clearly resolved into two distinct components. This also reveals a potential third source to the upper left of source \#1b. However, the signal-to-noise is low, and the size is considerably smaller than the synthesised beam. As such, we do not include this source in any analyses, but simply note that it is potentially another fragment, which could indicate that this is a multiple ($n$ \textgreater \ 2) system. 

\begin{table*}
\begin{center}
\label{tab:cores}
\caption{The below table displays the properties of the identified continuum sources. Shown are the cores identified, along with the enclosed flux, central coordinates (as Galactic coordinates, $l$ and $b$), radius (calculated by taking the exact area of each dendrogram leaf and determining the radius of a circle of equal area), core mass assuming a dust temperature of 22~K, number density ($n$, assuming spherical symmetry), and whether or not corresponding outflow emission was detected. All mass estimates assume a gas-to-dust ratio of 100 and a distance of 8.1~kpc \citep{Gravity19, Reid_2019}. $^{\dagger}$The masses reported for \#1a/b have been determined assuming a dust temperature of 22~K (upper mass limit), and assuming the gas temperature of 167/120~K, respectively, as determined from the CH$_{3}$CN emission (lower mass limit, see section \hyperref[sec:gas_temp]{3.6.1}). For all other sources, we do not detect any molecular lines that can be used to estimate gas temperatures.}
  \begin{tabular}{ccccccccc}
    \hline
    Source \# & Integrated flux	& $l$ & $b$ &	Radius & Mass & $n$ & Outflow\\
    & (mJy)	& ($^{\circ}$) & ($^{\circ}$) & (AU) & (\msun) & ($10^{7}$~\cmq) & detected?\\ \hline
    1 & 47.87 & 0.261034 & 0.0160561 & 4847 & 64.2 & 4.8 & Yes\\ \hline
    1a & 12.10 & 0.261008 & 0.016051 & 1300 & 1.7 - 16.2$^{\dagger}$ & 6.6 - 62.5$^{\dagger}$ & Yes \\
    1b & 14.34 & 0.261042 & 0.016058 & 1300 & 2.7 - 18.0$^{\dagger}$ & 10.4 - 69.5$^{\dagger}$ & Yes \\ \hline
    2 & 2.15 & 0.260737 & 0.0153934 & 1475 & 2.9 & 7.7 & Yes\\
    3 & 0.69 & 0.261774 & 0.0156472 & 1275 & 0.9 & 3.7 & No\\
    4 & 0.95 & 0.262961 & 0.0158849 & 1543 & 1.3 & 3.0 & Yes\\
    5 & 2.63 & 0.260413 & 0.0152248 & 1711 & 3.5 & 5.9 & Yes\\
    6 & 1.48 & 0.263037 & 0.0161058 & 1682 & 2.0 & 3.6 & No\\
    7 & 0.62 & 0.260976 & 0.0162781 & 1179 & 0.8 & 4.1 & Maybe\\
    8 & 0.41 & 0.262593 & 0.0163765 & 945 & 0.6 & 6.0 & Yes\\
    9 & 3.27 & 0.260611 & 0.016462 & 1687 & 4.4 & 7.8 & Yes\\
    10 & 1.61 & 0.262265 & 0.016906 & 1236 & 2.2 & 9.9 & No\\
    11 & 1.63 & 0.262183 & 0.0169606 & 1161 & 2.2 & 11.9 & Yes\\
    12 & 1.06 & 0.261871 & 0.0199851 & 945 & 1.4 & 14.1 & No\\
    13 & 1.68 & 0.261911 & 0.0200653 & 1139 & 2.3 & 13.2 & No\\
    14 & 7.17 & 0.26156 & 0.0208225 & 2613 & 9.6 & 4.6 & Yes\\
    15 & 2.34 & 0.262223 & 0.0201618 & 2206 & 3.1 & 2.4 & No\\
    16 & 1.34 & 0.262202 & 0.0203759 & 1568 & 1.8 & 4.0 & Yes\\
    17 & 0.76 & 0.257135 & 0.0163917 & 1302 & 1.0 & 3.8 & No\\
    \hline 
    \hline
  \end{tabular}
\end{center}
\end{table*}

\subsection{Molecular Line Data}\
\label{sec:lines}
 The full details of the spectral setup are given in Table \hyperref[tab:spws]{2}. Lines that were specifically targeted are SiO~(5-4) and $^{13}$CO~(2-1) as these are traditionally good outflow tracers \citep[e.g.][and references therein]{Bally_16}, 3 para-H$_{2}$CO transitions, which can be used to measure gas temperatures in the range \ap 50 -- 150~K, and the J=12-11 k-ladder of CH$_{3}$CN, which can be used to measure higher gas temperatures and is often found in the vicinity of protostars. 
 
Manual inspection of all spectral windows towards the continuum sources reveals that significant compact line emission is only detected towards the central sources, 1a and 1b. We do not find any single emission line in our spectral setup that can reliably trace all continuum sources. Such a lack of correspondence between continuum and molecular line emission has been noted previously in \brick, and in the CMZ in general \citep[e.g.][]{rathborne15, Jens_16a, Henshaw_2019}, though \citet{Barnes_2019} recently reported a suite of molecular lines at \ap 260~GHz which do reliably trace the continuum structure on 1\asec \ scales in the CMZ dust ridge clouds D, E, and F. We defer detailed analysis of the molecular line emission to a future publication, and focus only on the SiO, $^{13}$CO, and CH$_{3}$CN emission in the following sections.

\subsection{SiO (5-4) emission}

\begin{figure*}
\begin{center}
\label{fig:all_outflows}
\includegraphics[scale=0.75, angle=0]{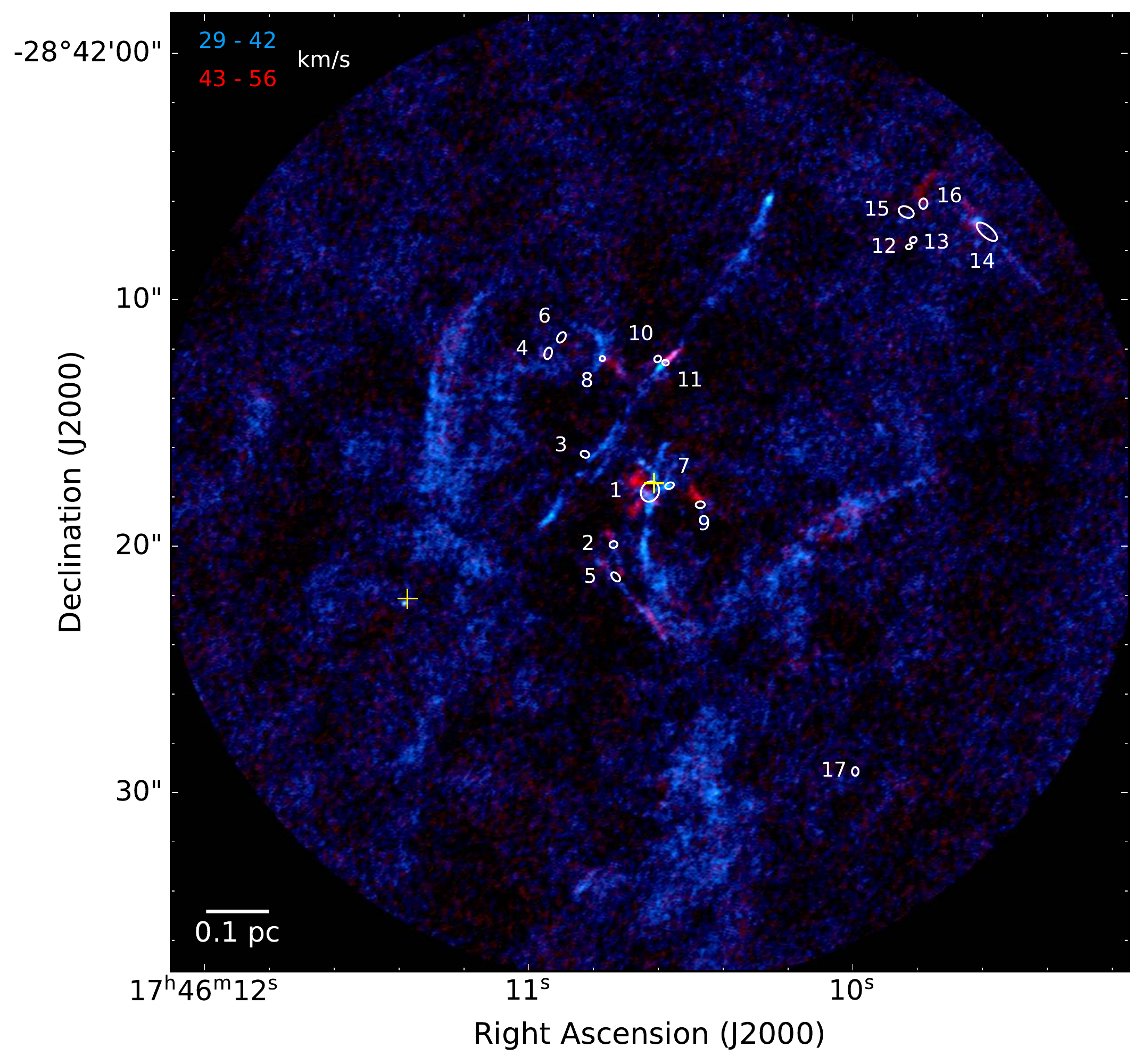}
 \caption{Two-colour image highlighting the outflows as traced by SiO (5-4) emission in our ALMA field. The blue-shifted emission is integrated between 29 -- 42~\kms, and the red-shifted emission 43 -- 56~\kms. Continuum sources are highlighted by white ellipses, the extent of which corresponds to the structures determined using dendrograms. Each continuum source is also numbered. The yellow crosses show the position of water masers from \citet{Lu19}.}
\end{center}
\end{figure*}

As discussed in section \hyperref[sec:intro]{1}, this region in \brick \ has been noted in the literature due to the presence of a bright, compact continuum source that is associated with water maser emission. While this is potentially indicative of active star formation, no definitive signatures have previously been found. To directly address the star forming nature of the source, we searched for outflows, as they are unambiguous signatures of active star formation \citep[e.g.][]{Bally_16}. We explicitly targeted the SiO (5-4) 217.105~GHz transition, as this is a well-established outflow tracer. Previous observations of \brick \ on larger scales with the SMA and ALMA did not detect any signatures of outflow emission in the cloud in SiO~(5-4), $^{12/13}$CO~(2-1) or any other molecular transitions \citep[e.g.][]{Brick_jens, Brick_Katharine, Brick_jill}. More generally, protostellar outflows have largely eluded detection in the CMZ. To-date, they have only been detected in the massive star-forming region Sagittarius B2 \citep{Qin08, Higuchi15} and a few high-mass CMZ clouds \citep{Lu2021}.

Figure \hyperref[fig:all_outflows]{4} shows a two-colour map, where the blue and red correspond to the integrated intensity of the SiO~(5-4) emission for the blue- and red-shifted emission across our ALMA field. The blue-shifted emission has been integrated over 29 -- 42~\kms, and the red-shifted emission over 43 -- 56~\kms. There is more compact and diffuse SiO emission detected at both lower and higher velocities (see Figures \hyperref[fig:SiO_appendix1]{9 -- 13}). The range displayed here has been chosen to highlight the outflows while minimising confusion from more diffuse emission.

We clearly detect multiple bipolar outflows associated with many of the continuum sources, along with larger-scale emission in the field. Thus, we directly confirm that active star formation is occurring in \brick. Overall we identify outflow signatures associated with sources \#1 (a \& b), \#2, \#4, \#5, \#8, \#9, \#11, \#14, and \#16. It is difficult to determine if source \#7 has any associated outflow emission, as its projected position is very close to the strong emission from source \#1. We do not detect any outflows towards sources \#3, \#6, \#10, \#12, \#13 or \#17. 

The most striking outflow is that associated with source \#11. Figure \hyperref[fig:all_outflows]{4} shows that this source has bright, compact emission in the blue- and red-shifted lobes of the outflow, along with fainter emission from a highly-collimated jet that extends \ap 0.7~pc across and is mostly blue-shifted. It is the brightest, most symmetrical, and most collimated outflow detected in this region. Sources \#10 and \#11 are very close together in projected separation, and are essentially identical in terms of their projected size, brightness, and mass (assuming equal temperatures). Despite this, only \#11 appears to be driving an (observable) outflow, which may indicate that they are in different evolutionary phases.

\begin{table*}
\centering
\label{tab:outflow_properties}
\caption{Properties of the outflows detected via SiO (5-4) emission determined for each of the blue- and red-shifted outflow lobes. Shown for each lobe is the projected size (l$_{\textrm{proj}}$), velocity range ($v_{\textrm{range}}$), peak intensity ($I_{\textrm{peak}}$), mean integrated SiO intensity ($\langle \int \textrm{T}_{\textrm{mb}}~dv \rangle$), mean SiO column density ($\langle \textrm{N}_{\textrm{SiO}} \rangle$), mass (M), momentum (P), kinetic energy (E$_{\textrm{k}}$), dynamical time ($\tau_{\textrm{dyn}}$, calculated by taking the full extent of the velocity range), and the mass outflow rate ($\dot{\textrm{M}}_{\textrm{out}}$). $^{\dagger}$A fractional SiO abundance of 1$\times$10$^{-8}$ is assumed. This is subject to large uncertainties, which are discussed in section \hyperref[sec:outflow_properties]{3.4}. $^{*}$The properties for source 11 are given both for the bright, compact outflow emission, as well as for the full flow including the faint extended emission, the latter of which are given in parentheses.}
\resizebox{\linewidth}{!}{%
  \begin{tabular}{cccccccccccc}
    \hline
    Source & Lobe & l$_{\textrm{proj}}$ & $v_{\textrm{range}}$ & $I_{\textrm{peak}}$ & $\langle \int \textrm{T}_{\textrm{mb}}~dv \rangle$ & $\langle \textrm{N}_{\textrm{SiO}} \rangle$ & M$^{\dagger}$ & P & E$_{\textrm{k}}$ & $\tau_{\textrm{dyn}}$ & $\dot{\textrm{M}}_{\textrm{out}}$\\
     $\#$ & & (10$^{4}$~AU) & (\kms) & (K) & (K \kms) & (10$^{13}$~cm$^{-2}$) & (M$_{\odot}$) & (M$_{\odot}$~km~s$^{-1}$) & (M$_{\odot}$~km$^{2}$~s$^{-2}$) & (10$^{3}$~yr) & (10$^{-5}$~M$_{\odot}$~yr$^{-1}$)\\ \hline
    \multirow{2}{*}{1} & Blue & 1.7 & [10, 34] & 5 & 38 & 5.1 & 0.12 & 2.2 & 28 & 3.4 & 3.7 \\ & Red & 2.3 & [34, 71] & 6 & 35 & 4.6 & 0.48 & 7.7 & 96 & 2.9 & 16.2\\ \hline
    \multirow{2}{*}{2} & Blue & 1.0 & [36, 45] & 6 & 11 & 1.5 & 0.02 & 0.1 & 0.3 & 5.2 & 0.4 \\ & Red & 0.5 & [40, 49] & 5 & 13 & 1.7 & 0.02 & 0.1 & 0.4 & 2.6 & 0.8 \\ \hline
    \multirow{2}{*}{4} & Blue & 0.2 & [8, 29] & 4 & 39 & 5.2 & 0.01 & 0.3 & 4 & 0.5 & 2.6\\ & Red & 0.2 & [36, 48] & 4 & 44 & 5.9 & 0.02 & 0.3 & 4 & 0.8 & 2.2\\ \hline
    \multirow{2}{*}{5} & Blue & 0.9 & [29, 39] & 4 & 12 & 1.5 & 0.04 & 0.2 & 0.5 & 4.3 & 0.5 \\ & Red & 2.8 & [37, 49] & 6 & 13 & 1.7 & 0.12 & 0.5 & 1.5 & 11.0 & 1.1 \\ \hline
    \multirow{2}{*}{8} & Blue & 0.5 & [15, 39] & 5 & 48 & 6.5 & 0.05 & 1.0 & 14 & 1.0 & 4.8\\ & Red & 0.7 & [47, 84] & 6 & 54 & 7.2 & 0.12 & 2.8 & 41 & 0.9 & 13.8\\ \hline
    \multirow{2}{*}{9} & Blue & 0.3 & [25, 34] & 3 & 26 & 3.4 & 0.02 & 0.4 & 5 & 1.6 & 1.1\\ & Red & 0.7 & [39, 58] & 5 & 35 & 4.8 & 0.06 & 1.1 & 14 & 1.8 & 3.7\\ \hline
    \multirow{2}{*}{11$^{*}$} & Blue & 0.7 (6.8) & [9, 36] & 10 & 62 (18) & 8.2 (2.3) & 0.12 (0.78) & 2.2 (8.9) & 25 (73) & 1.2 (11.9) & 10.3 (6.6)\\ & Red & 0.7 (6.8) & [36, 59] & 8 & 54 (22) & 7.3 (2.9) & 0.08 (1.08) & 1.0 (12.3) & 11 (100) & 1.4 (14.0) & 5.5 (7.7)\\ \hline 
    \hline
  \end{tabular}%
}
\end{table*}

At the centre of the field, the brightest continuum sources (\#1a \& \#1b) appear to be driving multiple outflows. While there is a clearly-defined bipolar outflow in the North-South direction (see also Figure \hyperref[fig:SiO_appendix1]{9}), there are several other components in different directions, particularly at higher velocities. This complex outflow structure may be due to the fact that there are at least two sources here, both of which could be driving outflows that may be interacting. It may also hint at a higher multiplicity, which is also potentially seen in the continuum emission (see Figure \hyperref[fig:robust_minus2]{3}).

In addition to the outflows, we also detect a number of features in the SiO emission. In particular, there is a significant amount of diffuse emission in the field. To the East of source \#1 there is diffuse arc of SiO emission (see Figures \hyperref[fig:all_outflows]{4}, \hyperref[fig:SiO_appendix1]{9b}, and \hyperref[fig:SiO_appendix4]{12a}). The origin of this emission is not clear. One possible interpretation is that it is tracing the shocked edge of a clump hosting the protostellar cluster, in which case this could imply that either the clump is moving supersonically towards the East, or that the lower density gas is being swept towards the West around the cluster.

Towards the South-West of the field, there is a significant amount of diffuse emission between \ap 5 -- 25~\kms \ (see Figures \hyperref[fig:SiO_appendix1]{9 -- 11}). We find that the SiO emission integrated over this velocity range spatially coincides with the edge of an arc-like structure that has previously been observed in \brick \ in a number of high-density/temperature and shock tracers, as well as CH$_{3}$OH masers \citep{Higuchi14, Mills15, Henshaw_2019}. The origin of this arcuate structure is unknown, but recent results suggest that it could be a feedback-driven shell due to embedded star formation (Henshaw et al. in prep.).

We also detect significant SiO emission at velocities that are considerably lower than that of the V$_{lsr}$ of the \q{maser core} (which is \ap 40~\kms). In the range (-)18 -- (+)10~\kms \ there are linear features that look outflow-like, as well as arcuate structures that could be tracing shock-fronts (see Figure \hyperref[fig:SiO_appendix2]{10}). Given the low velocities of these features, it's likely that they are associated with foreground material along the line-of-sight, rather than \brick.

\subsection{Outflow properties}
\label{sec:outflow_properties}
Having detected a population of molecular outflows that are associated with the continuum sources in \brick, we now estimate their general properties, assuming local thermodynamic equilibrium and optically thin SiO emission. We also assume that the outflow emission is parallel to the plane of the sky, since we do not have any knowledge of possible inclinations. Though we clearly detect outflows associated with sources \#14 and \#16, we do not report their properties as they are very close to the edge of the field where the sensitivity is worse and the noise is higher due to the primary beam response. The outflow properties for the remaining blue- and red-shifted pairs of lobes are presented in Table \hyperref[tab:outflow_properties]{4}.

The projected size (l$_{\textrm{proj}}$) of each outflow lobe is calculated assuming a distance of 8.1~kpc, with typical sizes of 10$^{3-4}$~AU. Combining these sizes with the full extent of the measured velocity ranges, we measure dynamical timescales ($\tau_{\textrm{dyn}}$) of \ap 10$^{3-4}$~years. To estimate the column density of the outflow emission we follow the formalism presented in section 3.4 of \citet{Li_19}. An excitation temperature of 30~K is assumed. The column density is not too strongly dependent on the excitation temperature. An increase from 30~K to 200~K, which covers the range of measured temperatures for the bulk of the gas in \brick \ \citep{Adam_cmz_temp, Immer16, Krieger17}, only increases the estimated column density by a factor of \ap 2. Under these assumptions, typical mean column densities of \ap 10$^{13-14}$~cm$^{-2}$ are measured for the outflows in \brick.

To determine outflow masses, we sum the emission over the spatial extent of each lobe and along the velocity axis over the relevant channels, and estimate the outflow mass as:
\begin{equation}
\label{eq:outflow_mass}
M_{\textrm{outflow}} = \sum_{A,v} X^{-1}_{\textrm{SiO}} N_{\textrm{SiO}} A_{\textrm{pix}} \mu m_{\textrm{H}}
\end{equation}
\noindent
where $N_{\textrm{SiO}}$ is the SiO column density, $A_{\textrm{pix}}$ is the pixel area, $\mu$ is the mean molecular weight which is assumed to be 2.8 \citep{Jens_mass_in}, $m_{\textrm{H}}$ is the mass of hydrogen, and $X_{\textrm{SiO}}$ is the fractional abundance of SiO. The SiO abundance is subject to high uncertainty, with several orders of magnitude of spread reported in the literature. Estimates towards IRDCs report an abundance of 5$\times$10$^{-10}$ \citep{Sanhueza_12}, and measurements of some CMZ clouds report 6$\times$10$^{-10}$ \citep{Tsuboi_15}. \citet{Li_19} report average abundances in Galactic massive star-forming regions of 4$\times$10$^{-11}$, but with a scatter of \ap 2 orders of magnitude. \citet{SanchezMonge13} and \citet{Leurini114} find a range of \ap 1$\times$10$^{-9}$ to 3$\times$10$^{-8}$ for SiO outflows in massive star-forming regions. The best constraints towards \brick \ report an SiO abundance of 10$^{-9}$ from measurements on 26\asec \ scales \citep{Martin-Pintado97}. As these scales are significantly larger than those probed in this work, it is not clear whether this measurement should hold here. Most recently, \citet{Lu2021} estimate the SiO abundances in a sample of 43 outflows in a few CMZ clouds to be between $10^{-10}$ and $10^{-8}$, with a mean value of $2\times10^{-9}$ on scales of \ap 0.2\asec. However, the uncertainty in these abundances is at least one order of magnitude, and \brick \ was not included in their sample.

As we do not have any direct constraints on the SiO abundance in \brick \ on the spatial scales discussed in this paper, we assume an abundance of 10$^{-8}$ as a soft upper limit. This is consistent with upper limits measured in star-forming regions both in the Galactic disc and in the CMZ. This assumption means that any masses and dependent properties reported are considered to be likely lower limits. However, given the already-high abundance of gas-phase SiO in the CMZ, which could be enhanced even further in the vicinity of protostellar outflows due to high-velocity shocks \citep[e.g.][]{Schilke97}, it is plausible that the abundance may even be as high as a few $\times$10$^{-7}$ (Gusdorf et al., private communication). Such high SiO abundances have been assumed in the extreme star-forming region W51 \citep{Goddi_outflows_W51}.

Under these assumptions, we obtain outflow masses of order 10$^{-2}$ to 1~\msun, and mass outflow rates ($\dot{\textrm{M}}_{\textrm{out}}$) of 10$^{-6}$ to 10$^{-4}$~\msun~yr$^{-1}$. We also estimate the outflow momentum and kinetic energy as:
\begin{equation}
\label{eq:outflow_momentum}
P_{\textrm{outflow}} = \sum_{A,v} M \lvert v - v_{\textrm{lsr}} \rvert
\end{equation}

\begin{equation}
\label{eq:outflow_energy}
E_{\textrm{\textrm{k}, outflow}} = \frac{1}{2} \sum_{A,v} M \lvert v - v_{\textrm{lsr}} \rvert^{2}
\end{equation}
\noindent
where $v$ is the velocity at a given channel, and $v_{\textrm{lsr}}$ is taken to be the velocity of the source driving the outflow. In the majority of cases, the source velocity is not known due to the lack of line emission associated with most of the continuum sources (see section \hyperref[sec:lines]{3.2}). Thus, for most of the outflows, we assume the velocity of the central source, which is \ap 40~\kms. The estimated momenta range \ap 0.1 -- 12 \msun~\kms. and energies range \ap 0.5 -- 100~\msun~km$^{2}$~s$^{-2}$ (1$\times$10$^{44}$ -- 2$\times$10$^{45}$ erg) per lobe.

\subsubsection{Outflow energetics}
\label{sec:outflow_energy}

The total estimated energy contained in the detected outflows is \ap 8$\times$10$^{45}$ erg. To investigate what impact these outflows may have on the local environment, we first estimate the gravitational energy of the material in our full ALMA field. Using the dust column density map from Hi-GAL \citep[30\asec,][]{higal, Cara_higal, Mills17}, we estimate this mass to be \ap 7$\times$10$^{3}$~\msun \ within a radius of 19\asec. This corresponds to a gravitational energy of 5$\times$10$^{48}$ erg via $\textrm{E}_{\textrm{grav}}$ = GM$^{2}$/R. We also estimate the turbulent energy in the region as $\textrm{E}_{\textrm{turb}}$ = M($\sqrt{3}\sigma_{\textrm{los,1D}}$)$^{2}$/2, where $\sigma_{\textrm{los,1D}}$ is the one dimensional line-of-sight velocity dispersion. To measure this velocity dispersion, we take an averaged spectrum across our ALMA field using the HNCO emission from the MALT90 survey \citep{MALT90_1}. The angular resolution of the MALT90 data is 38\asec, which is approximately the same size as our ALMA field of view. The averaged HNCO spectrum shows two overlapping velocity components. Fitting the brightest component with a single Gaussian yields $\sigma_{\textrm{los,1D}}$ \ap 4.3~\kms. If we include both components, this increases to \ap 12~\kms. Using this range we estimate the turbulent energy to be \ap 4$\times$10$^{48}$ -- 3$\times$10$^{49}$ erg.

These results suggest that the detected population of SiO outflows are not energetic enough to drive the local turbulence or to significantly disrupt the local material. We reiterate that the measured outflow masses and energies are potential lower limits and so the impact of the outflows could be larger. However, the fractional SiO abundance would need to be 3 -- 4 orders of magnitude lower in order for the outflow energy to be similar to the gravitational and turbulent energy, which seems unlikely given the high abundance of SiO in the gas phase the CMZ in general \citep[\ap 10$^{-9}$, e.g.][]{Martin-Pintado97, Amo-Baladron09}.

\subsection{$^{13}$CO (2-1) emission}

CO is the most commonly used tracer for identifying outflows, due to the high abundance of the molecule along with the relatively low energies of the lower rotational states \citep[e.g.][and references therein]{Bally_16}. We therefore searched for outflow emission via the $^{13}$CO~(2-1) transition ($^{12}$CO was not covered in our spectral setup). However, the $^{13}$CO emission towards \brick \ is complex and widespread on large spatial scales, thus making it difficult to isolate any emission potentially associated with outflows. Despite this, we do detect a significant amount of interesting structure via the $^{13}$CO emission, which is discussed below.

\begin{figure*}
\begin{subfigure}[b]{0.49\textwidth} 
    \includegraphics[width=\linewidth]{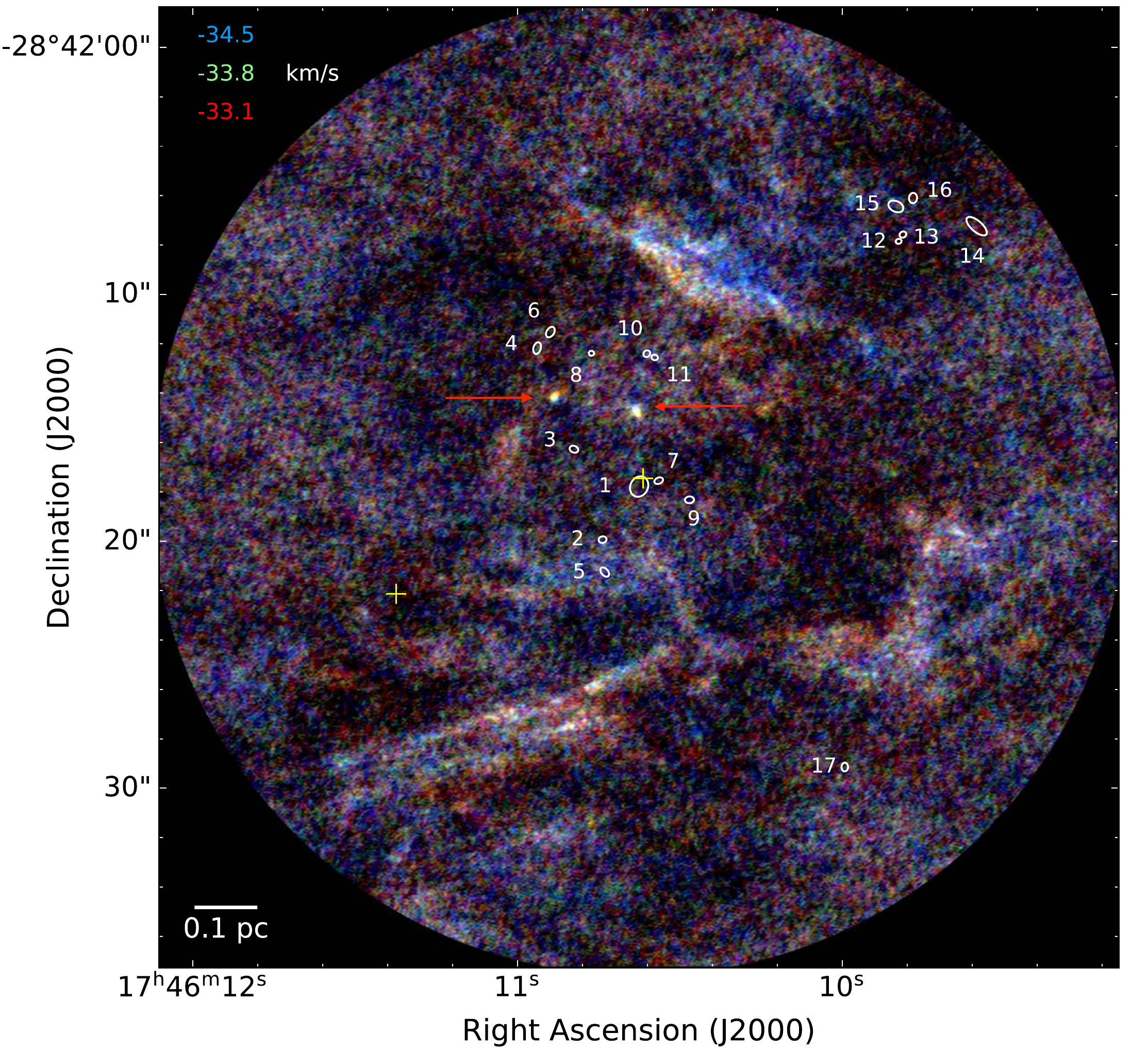}
    \caption{} 
\end{subfigure}%
\hfill
\begin{subfigure}[b]{0.49\textwidth}
    \centering
    \includegraphics[width=\linewidth]{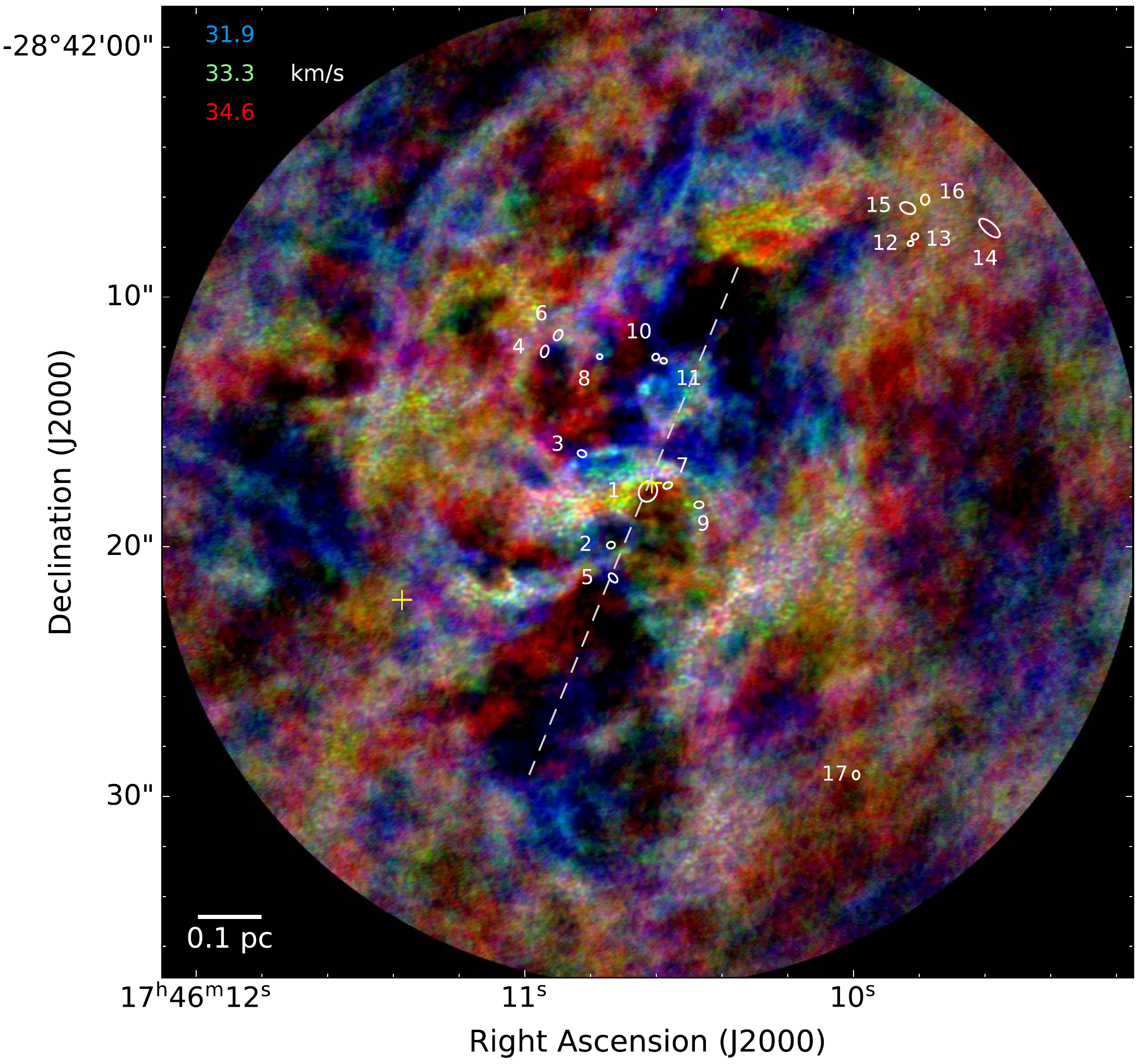}
    \caption{} 
    \end{subfigure}

\begin{subfigure}[b]{0.49\textwidth} 
    \includegraphics[width=\linewidth]{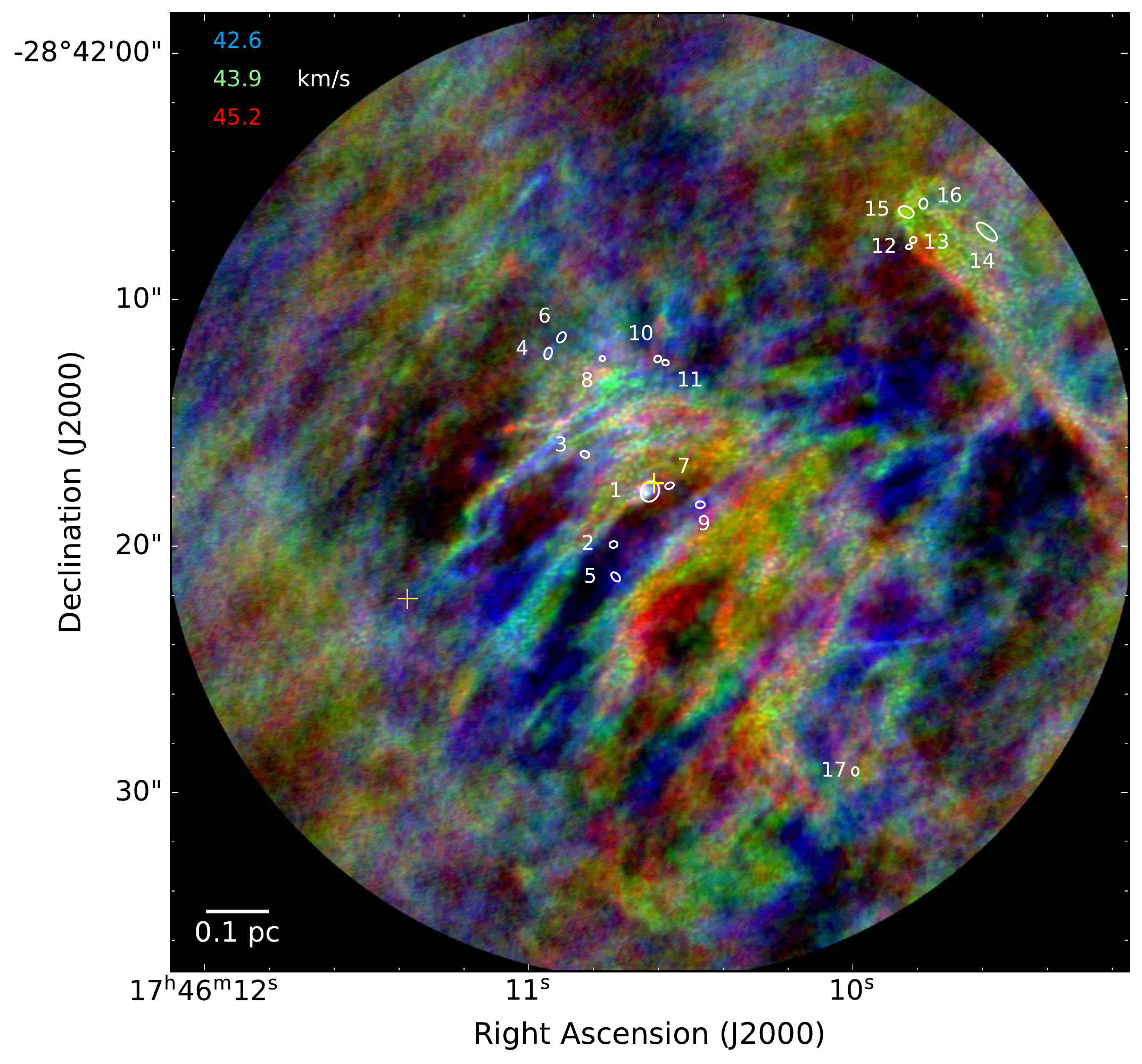}
    \caption{} 
\end{subfigure}%
\hfill
\begin{subfigure}[b]{0.49\textwidth} 
    \includegraphics[width=\linewidth]{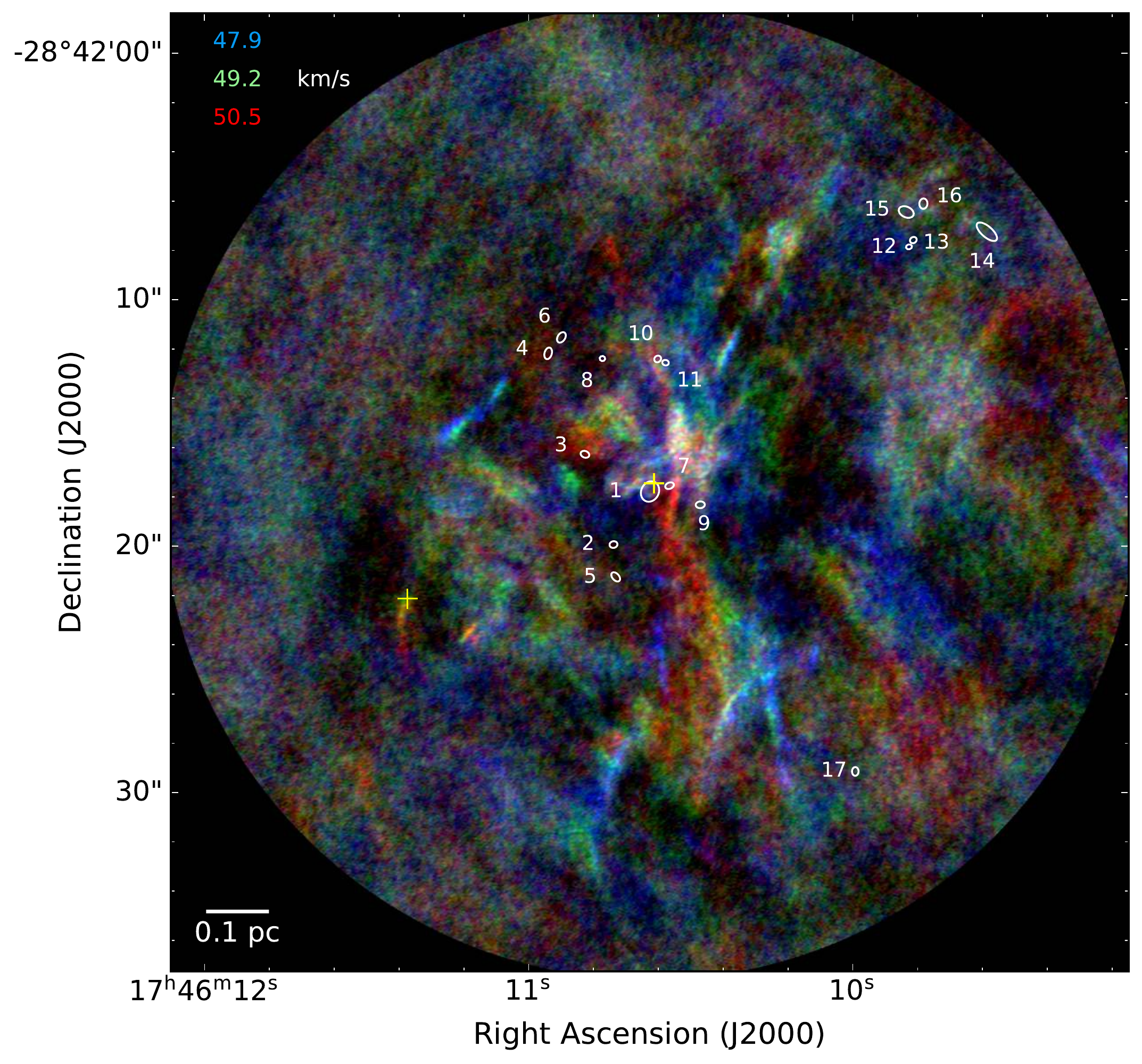}
    \caption{} 
\end{subfigure}%
    \caption{Three-colour figures showing the $^{13}$CO emission at different velocities. \textbf{(a)}: -34.5 (blue), -33.8 (green), and -33.1 (red) \kms, \textbf{(b)}: 31.9 (blue), 33.3 (green), and 34.6 (red) \kms, \textbf{(c)}: 42.6 (blue), 43.9 (green), and 45.2 (red) \kms, \textbf{(d)}: 47.9 (blue), 49.2 (green), and 50.5 (red) \kms. Continuum sources are highlighted by white ellipses, the extent of which corresponds to the structures determined using dendrograms. Each continuum source is also numbered. The yellow crosses show the position of water masers from \citet{Lu19}. The red arrows in \textbf{(a)} highlight two bright, compact knots of $^{13}$CO emission. The grey dashed line in \textbf{(b)} shows the axis of symmetry of a cavity-like region.} 
\label{fig:13CO_emission_maps}
\end{figure*}

\begin{figure*}
\begin{subfigure}[b]{0.49\textwidth} 
    \includegraphics[width=\linewidth]{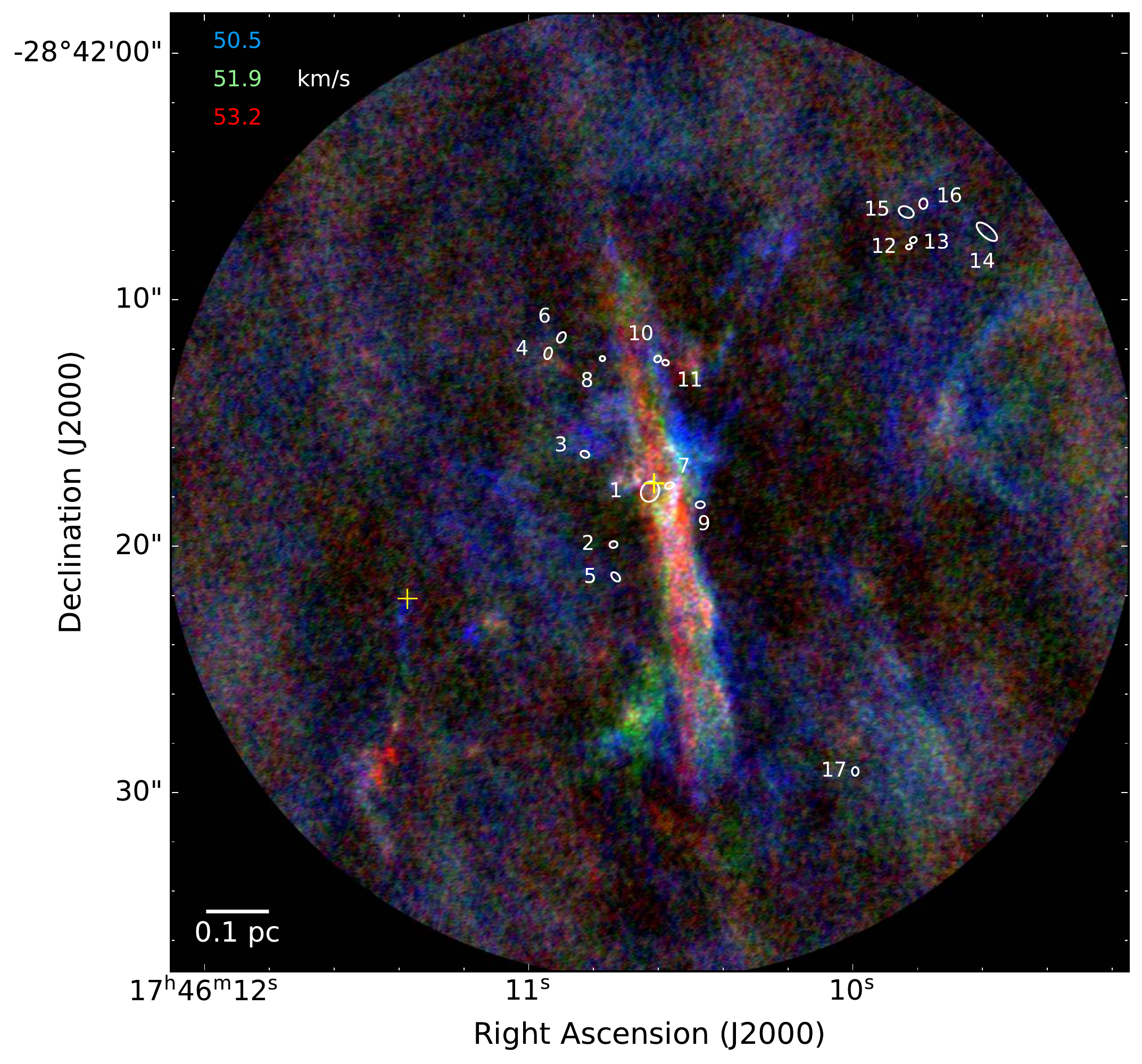}
    \caption{} 
\end{subfigure}%
\hfill
\begin{subfigure}[b]{0.49\textwidth}
    \centering
    \includegraphics[width=\linewidth]{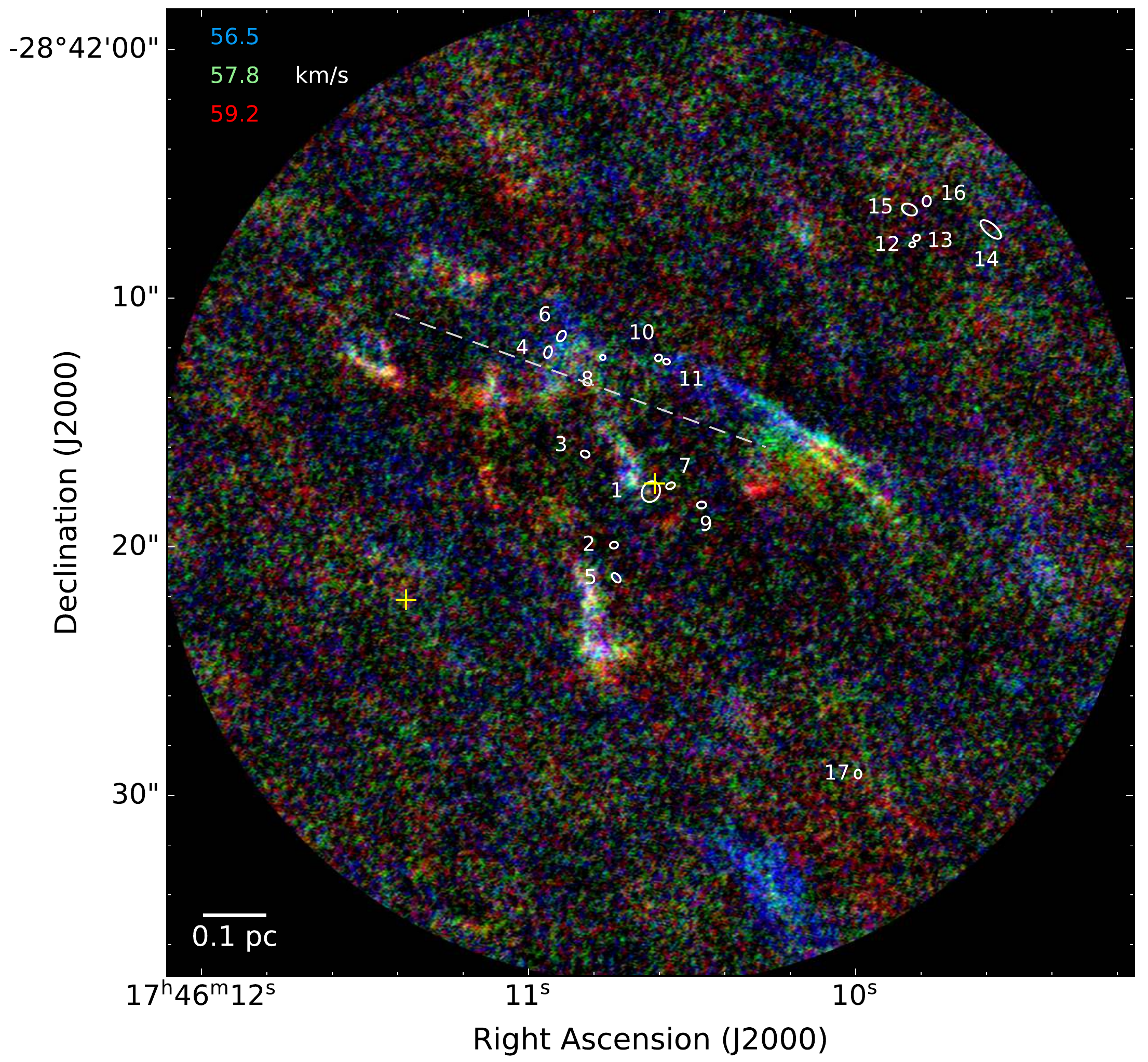}
    \caption{} 
    \end{subfigure}
    \caption{As in Figure 5. Three-colour figures showing the $^{13}$CO emission at different velocities. \textbf{(a)}: 50.5 (blue), 51.9 (green), and 53.2 (red), \textbf{(b)}: 56.5 (blue), 57.8 (green), and 59.2 (red) \kms. The grey dashed line in \textbf{(b)} shows the axis of symmetry of an hourglass-shaped region that may be tracing an outflow cavity from source 8.} 
\label{fig:13CO_emission_maps_2}
\end{figure*}

Unlike with the SiO emission, we do not detect clear-cut outflow emission in $^{13}$CO. Though we do identify a large number of linear features, particularly towards the central region of the field, that may be associated with outflows and/or outflow cavities (see Figure \hyperref[fig:13CO_emission_maps]{5d}). As shown in Figure \hyperref[fig:13CO_emission_maps]{5b}, we also identify large regions lacking in emission that are roughly centred on source \#1. This could be tracing a pair of outflow cavities, each of which is \ap 10\asec \ in extent. We also see an hourglass-shaped emission structure between \ap 56 -- 60~\kms \ (see Figure \hyperref[fig:13CO_emission_maps_2]{6b}). This structure is centred on source \#8, the red-shifted SiO outflow of which is along the axis of symmetry of the hourglass, suggesting that this structure may be tracing a cavity that has been shaped by the outflow from source \#8.

One particularly interesting feature is the \ap 50~\kms \ tail of emission that is associated with the water maser to the South-East of the centre of the field, as shown in Figure \hyperref[fig:13CO_emission_maps]{5d}. This $^{13}$CO tail also coincides with a bright, compact knot of SiO emission that is very close to the second water maser (see Figure \hyperref[fig:all_outflows]{4}). This water maser was reported in \citet{Lu19}, where they note that the maser emission was not found to coincide with any continuum source on \ap 0.2\asec \ scales. Our observations confirm this on \ap 1000~AU scales -- we do not detect any continuum source at this location. However, the detection of both SiO and $^{13}$CO at this location, combined with the presence of a water maser, suggest a protostellar nature. The fact that we do not detect any continuum emission could imply that there truly is no source at this location, or if there is a source, then it is below our detection limit.

At velocities \ap 52~\kms, there is very linear region of $^{13}$CO emission that is centred on source \#1 and spans \textgreater \ 1~pc with a position angle of \ap 15 degrees (see Figure \hyperref[fig:13CO_emission_maps_2]{6a}). This emission is constrained to just a few \kms, and shows a \q{braided} structure, which appears to rotate when stepping through in velocity. Such a structure may plausibly be caused by precession of the central binary source shaping the outflow emission \citep[e.g.][]{fendt98}, though it is not clear that the origin of this emission is due to outflowing material. 
2
As shown in Figure \hyperref[fig:13CO_emission_maps]{5c}, at the approximate V$_{lsr}$ of the central cluster of continuum sources (\ap 41~\kms), the $^{13}$CO emission is complex. Most notably there are a series of arc-like filamentary structures with a roughly South-East -- North-West orientation. This \q{bear-claw} like structure is reminiscent of the HCO$^{+}$ absorption filaments observed in \brick \ on larger scales \citep{Bally14}. Inspection of the HCO$^{+}$ cube reveals that the two prominent broad-line absorption filaments shown in Figure 1 of \citet{Bally14} directly cross the centre of our ALMA field and coincide with the arcuate $^{13}$CO structures, both in position and velocity. In Figure \hyperref[fig:13CO_emission_maps]{5c}, we also see some interesting structure at the location of the sources \#12 -- \#16 towards the North-West edge of the field. This cluster of sources appears to be at the tip of a cometary-like structure, the trailing edge of which is red-shifted.

At velocities that are significantly blue-shifted with respect to the cloud (\ap -35~\kms) we also see some emission. In particular we identify two bright, compact knots of $^{13}$CO emission just North of the central continuum sources (see Figure \hyperref[fig:13CO_emission_maps]{5a}). These knots are confined to only a few \kms, but they display velocity gradients across their small extent. Given their location in the central cluster, these could be bullets or post-shock clumps due to protostellar outflows in the region. Though their highly blue-shifted velocities may suggest that they could be due to some unrelated emission along the line-of-sight.

Though there is a diverse amount of structure observed via the $^{13}$CO emission, we stress that caution must be taken when interpreting the data. The emission is complex and much of it is diffuse. As our observations do not utilise the 7~m or Total Power arrays of ALMA, we are missing short spacing data and so much of the larger-scale diffuse emission will be filtered out by the long baselines of the interferometer. Thus, while we can speculate on the origin of the emission, we do not present any strong conclusions based on the $^{13}$CO data.

\subsection{CH$_{3}$CN (12-11) emission}
\begin{figure*}
\begin{center}
\label{fig:ch3cn_moments}
\includegraphics[scale=0.68, angle=0]{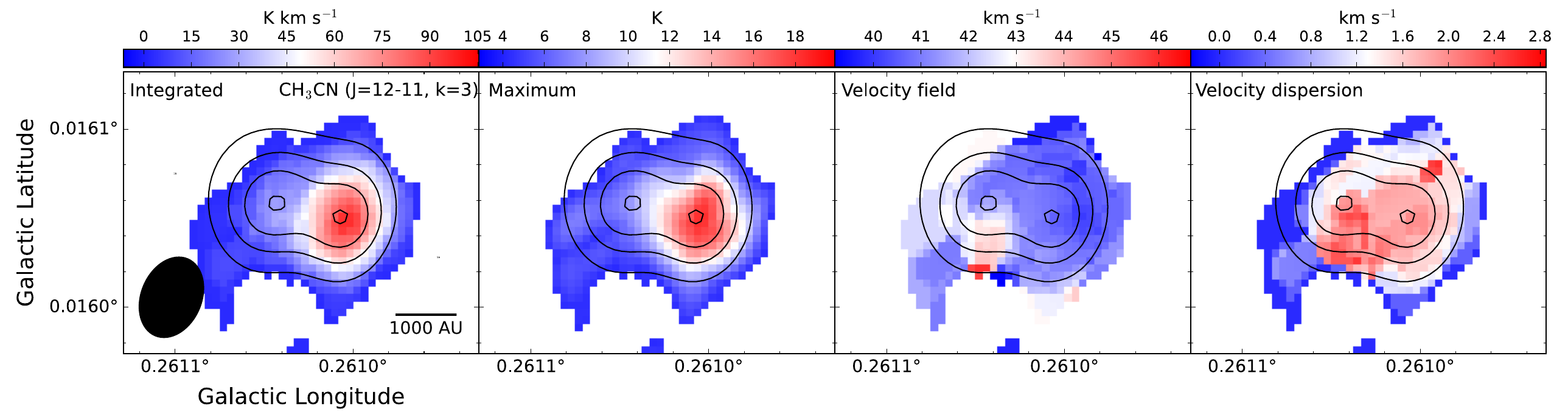}
 \caption{Moment maps of the CH$_{3}$CN J=12-11 k=3 emission towards the central sources (cores \#1a \& \#1b). The colourscale in each panel shows (from left to right) the integrated intensity, maximum intensity, velocity field, and velocity dispersion maps. Black contours show the continuum emission. While both cores are similarly bright in continuum emission, this shows that the CH$_{3}$CN emission is strongly dominated by core 1a. Contours are at the same levels as those described for Figure \hyperref[fig:full_continuum]{1c}.}
\end{center}
\end{figure*}

\begin{figure}
\begin{center}
\label{fig:ch3cn_fit}
\includegraphics[scale=0.28, angle=0]{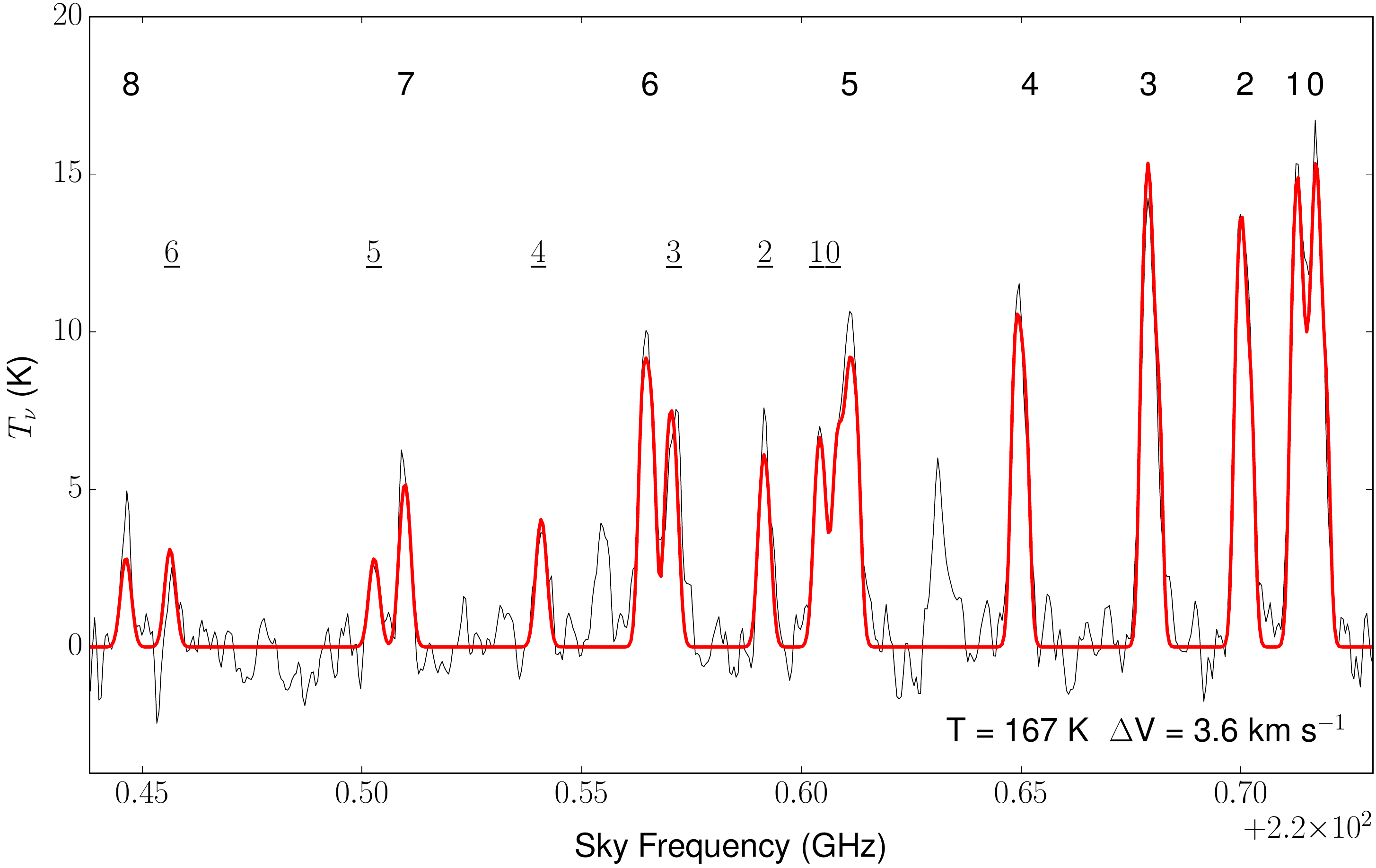}
 \caption{Beam-averaged spectrum of the CH$_{3}$CN J=12-11 emission towards the central source (core \#1a). The real emission is displayed in black, while the best-fitting model is overlaid in red. The bold numbers at the top of the plot indicate the k=0-8 components of CH$_{3}$CN, and the underlined numbers correspond to the k=0-6 components of isotopologue CH$_{3}$$^{13}$CN. The best-fitted temperature and line-width are shown at the bottom right. There are a few lines that are not fitted here, as they are not from CH$_{3}$CN or its isotopologues.}
\end{center}
\end{figure}

CH$_{3}$CN (methyl cyanide) is commonly used to trace small-scale gas kinematics towards hot protostellar cores, and the relative ratios of the k-components can be used to estimate gas temperatures and column densities \citep[e.g. some recent results include:][]{Henrik17,Ilee18,Maud18}. With this in mind, we targeted the J=12-11 k-ladder of CH$_{3}$CN at \ap 221~GHz as another means of identifying signs of star formation via locally-heated gas.

We detect CH$_{3}$CN only towards the central cores (sources \#1a \& \#1b). Figure \hyperref[fig:ch3cn_moments]{7} shows moment maps of the k=3 component emission towards these sources. While the two sources are similarly bright in continuum emission, the integrated and maximum intensity maps show that the CH$_{3}$CN emission is dominated by source \#1a. This may suggest that it is hotter and/or denser than \#1b, or it may be indicative of differing abundances or evolutionary phases.

Figure \hyperref[fig:ch3cn_fit]{8} shows a beam-averaged spectrum of the CH$_{3}$CN emission towards \#1a. The k=0-8 components of CH$_{3}$CN and k=0-6 components of the isotopologue CH$_{3}^{13}$CN are clearly detected towards the core. The upper energy levels of the CH$_{3}$CN k=0-8 components are 69, 76, 97, 133, 183, 247, 326, 419, and 525~K. Their detection therefore indicates that the gas is hot, and is likely internally heated by an embedded protostar(s).

\subsubsection{Gas temperature modelling}
\label{sec:gas_temp}
Measuring gas temperatures to be used as a proxy for the dust temperature is a fairly common practice in star formation studies. However, measurements in the CMZ have shown that the gas and dust are thermally decoupled on large scales (\ap 30\asec), with gas temperatures that are often many factors greater than measured dust temperatures \citep{Adam_cmz_temp, Immer16}. As we begin to probe sub-parsec scales in the CMZ, we no longer have dust temperature measurements, as these scales are not accessible with facilities such as Herschel or SOFIA at the distance of the CMZ. This leaves a large uncertainty in the dust temperatures, and hence mass estimates. However, SPH models of \brick \ suggest that the gas and dust should be close to thermalised in the density regime of the sources that we detect with our ALMA observations \citep[\ap 10$^{7-8}$~\cmq,][]{Clark_brick}. Thus, in the following we estimate gas temperatures (where possible) and use these as a proxy for the dust temperatures, while acknowledging the caveat that they may still be weakly decoupled. Any uncertainties in reported mass estimates are likely dominated by this.

We use the eXtended CASA Line Analysis Software Suite \citep[XCLASS\footnote{\url{https://xclass.astro.uni-koeln.de/}},][]{xclass} software package within CASA to simultaneously fit the CH$_{3}$CN and CH$_{3}^{13}$CN emission to obtain an estimate of the average gas temperature towards sources \#1a and \#1b. We opt to fit a single average spectrum rather than a pixel-by-pixel fit, as the source is not well sampled by the synthesised beam. We assume a filling factor of unity.

The resulting single-component fit for \#1a is shown in Figure \hyperref[fig:ch3cn_fit]{8}. The best-fitting parameters for the gas temperature, column density, and line-width are 167~K, 8.9$\times$10$^{16}$~cm$^{-2}$, and 3.6~\kms, respectively. This temperature is likely greater on smaller scales, close to the protostar(s), where the relative intensities of the higher k-components are likely to be greater. Note that there are several peaks that are not fit by the model. This is because those emission lines are not associated with CH$_{3}$CN/CH$_{3}^{13}$CN .

If we take this estimated gas temperature of 167~K and assume that the gas and dust are thermalised at these densities, we can use this to better constrain the lower limit of the dust mass. In section \hyperref[sec:continuum]{3.1} we estimated an upper mass limit for source \#1a of 16~\msun \ at 22~K. Assuming a dust temperature of 167~K, this mass estimate decreases to \ap 2~\msun. We also estimate the average gas temperature of \#1b to be 120~K, using the same method. Using this as the dust temperature, the mass estimate decreases from 18~\msun \ to 3~\msun. 

\subsubsection{Dynamical masses of the central sources 1a \& b}
\label{sec:dynamical_mass}

If this system is a protostellar binary, then another method of constraining their masses is through a dynamical argument. Looking at the two right-most panels in Figure \hyperref[fig:ch3cn_moments]{7} -- the 1$^{\textrm{st}}$ and 2$^{\textrm{nd}}$ moments -- there is a small difference between the velocities and velocity dispersions of the two sources. If we assume that the sources are of equal mass, and that their measured velocities are the maximum line-of-sight velocities (i.e. that we are viewing the system edge-on), then we can use a simple gravitational-kinetic energy balance argument to estimate their masses. For a binary separation of 1150~AU, and a velocity difference between the sources of 1.5~\kms, we estimate that the dynamical mass of each source is \ap 1.5~\msun. If we now consider the possible inclination, which is observationally unconstrained, this calculation is modulated by a \emph{sin(i)} term. For 15$^{\circ}$ $\leq$  \emph{i} $\leq$ 90$^{\circ}$, the estimated dynamical mass range is 5.6~\msun \ $\geq$ M$_{\textrm{dyn}}$ $\geq$ 1.5~\msun. 

While this is a simple argument with several caveats, and one that requires that the sources are actually in a binary system, it demonstrates that the dynamical masses in such a scenario are consistent with the dust masses that have been estimated assuming that T$_{\textrm{dust}}$ = T$_{\textrm{gas}}$. This suggests that this assumption may be valid, and that the dust and gas may be thermalised at these densities (\ap 10$^{7-8}$~\cmq) in the CMZ. If instead we assume the upper mass limit for these sources of \ap 18~\msun \ (for T$_{\textrm{dust}}$ = 22~K), this would require a velocity difference between them of 3 -- 5.5 \kms \ for 15$^{\circ}$ $\leq$  \emph{i} $\leq$ 90$^{\circ}$, which is several factors greater than what is observed.

\section{Discussion}
\label{sec:discussion}
\subsection{Fragmentation}
Given the extreme conditions in the CMZ, particularly the elevated gas temperatures and high gas densities \citep[e.g][]{Adam_cmz_temp, Immer16, Krieger17, Mills18}, it is pertinent to investigate the continuum structure and the nature of the fragmentation that occurs in the molecular clouds here. The thermal Jeans mass is given by:
\begin{equation}
\label{eq:jeans_mass}
M_J = \frac{\pi^{5/2} c_s^{3}}{6\sqrt{G^3\rho}}
\end{equation}

\noindent where $c_s = (k_bT/\mu m_\textrm{H})^{1/2}$ is the sound speed, G is the gravitational constant, and $\rho$ is the volume density. For the \q{maser core} on 1\asec \ scales, \citet{Jill_pdf_2014} estimate a range of densities from 1 -- 3$\times$10$^{6}$~\cmq, for a dust temperature range of 50 -- 20~K, respectively. Taking this range of parameters, the resulting thermal Jeans mass ranges from 0.35 -- 2.15~\msun. This range is broadly consistent with the masses that we estimate for the 1.3~mm continuum sources in our data, for which the median mass is 2~\msun. We reiterate that our mass estimates may be affected by significant uncertainties, primarily due to spatial filtering and unconstrained dust temperatures on these scales. Nonetheless, we find that the observed structure on small spatial scales is generally consistent with thermal Jeans fragmentation within the uncertainties.

{While all of the individual core masses and the median mass are consistent with thermal Jeans fragmentation, the full extent of the central source \#1 significantly exceeds the thermal Jeans mass. Subtracting the flux contributed by the embedded sources \#1a \& b, the enveloping mass is \ap 35~\msun \ (assuming T$_{\textrm{dust}}$ = 22~K), which is 16 -- 100 times greater than the thermal Jeans mass. This suggests that thermal fragmentation alone is not sufficient to explain the properties of core \#1, and that turbulent and/or magnetic support may contribute more significantly \citep[e.g.][]{Zhang09, Wang14}.

To investigate this, we can substitute the sound speed ($c_s$) in Equation \hyperref[eq:jeans_mass]{5} with the velocity dispersion of the gas ($\sigma$), under the assumption that the total velocity dispersion is a suitable proxy for the turbulent linewidth \citep[e.g.][]{Wang14, Li_19}. As explained in Section \hyperref[sec:outflow_energy]{3.4.1}, the velocity dispersion measured with single-dish data for this region of the cloud is in the range of 4 -- 12~\kms. Substituting this range in place of the sound speed yields turbulent fragmentation masses of thousands of~\msun \ for the range of densities given. This is significantly larger than the masses of the fragments that we observe. Even if we assume the small-scale velocity dispersion as measured towards the central sources via the CH$_{3}$CN emission ($\Delta$V = 3.6~\kms, $\sigma$ = 1.5~\kms, see Section \hyperref[sec:gas_temp]{3.6.1}), the turbulent fragmentation mass is \ap 50 -- 100~\msun. This further supports the conclusion that the thermal pressure is likely dominating the fragmentation process in this region of \brick, not the turbulent pressure.

We also consider the thermal Jeans length, given by: 
\begin{equation}
\lambda_J = c_s\sqrt{\frac{\pi}{G\rho}}
\end{equation}

Taking the same range of densities and temperatures as for the Jeans mass estimation, we obtain a range of 0.8 -- 1.4$\times$10$^{4}$~AU for the thermal Jeans length. Using a nearest neighbour algorithm, we find that the mean projected separation of the continuum sources in our ALMA field is 4.5$\times$10$^{4}$~AU. If we restrict this to only consider the sources in the central region of the field (sources 1 -- 11), this reduces slightly to 3.4$\times$10$^{4}$~AU. This is generally consistent with the expected separations from thermal Jeans fragmentation, though the observed separations are larger by a factor of a few. 

These results suggest that even in a cloud that is so turbulent \citep{Henshaw_2019, Henshaw_2020} and in such an extreme environment as the CMZ, thermal Jeans fragmentation may still dominate the fragmentation properties on protostellar scales. Results outside of the CMZ conclude that thermal Jeans fragmentation is sufficient to explain the observed core properties in a variety of Galactic disc environments on similar scales \citep[e.g.][]{Alves07, Lada08, Beuther18_CORE}.

While we cannot draw definitive conclusions from a small number of cores in a region of a single cloud, this result is important. It suggests that while large-scale gas properties in the CMZ are strongly shaped by turbulence \citep[e.g.][]{Adam_cmz_temp, Henshaw_cmz}, the small-scale properties may be less sensitive to this, in which case star formation may proceed \q{normally} once it is underway. \citet{Lu19} recently came to a similar conclusion when comparing star formation efficiencies on large and small scales in CMZ clouds. Using ALMA data similar in resolution and sensitivity to those presented here, \citet{Lu20} also concluded that the structure in four other massive CMZ molecular clouds is consistent with thermal Jeans fragmentation. Ultimately, these results suggest that the process of star formation on protostellar scales in this Galactic environment is not fundamentally different, other than the fact that the initial fragmentation towards protostellar cores is inhibited below a higher critical density threshold when compared to the Galactic disc \citep[e.g.][]{Ginsburg18, Walker18, Barnes_2019}.

\subsection{There are no high-mass protostars in \brick's \q{maser core} ... \emph{yet}}
\label{sec:no_hmcores}

\brick \ is one of the most massive (\textgreater \ 10$^{5}$~\msun) and dense (\textgreater \ 10$^{4}$~\cmq) molecular clouds known to exist in the Galaxy that appears to be largely quiescent \citep[e.g.][]{Immer, Brick_jens, Brick_Katharine}. Given these properties, it has been proposed to represent an ideal candidate precursor to a massive stellar cluster \citep{Brick_snl, Brick_jill, Walker15, Walker16}. Indeed, some of the most extreme star clusters exist in this region of the Galaxy, such as the Arches and Quintuplet, with the former being the most dense young cluster in the Galaxy \citep{Arches_esp}. As these clusters are relatively young \citep[\ap 3.5 and 4.8~Myr, respectively,][]{Schneider14}, they had to have formed in the CMZ. Comparison between the positions and motions of the Arches and Quintuplet with orbital models of the gas in the CMZ suggests that these clusters are consistent with having formed in the same region of the CMZ as \brick \ \citep{Diederik_orbit}.

Given the large reservoir of dense material in the cloud, along with its potential to form a massive stellar cluster, \brick \ is also an ideal region in which to search for precursors to high-mass stars. The most obvious region to search for such precursors is in the \q{maser core}, as it is the brightest and most compact continuum source detected within the cloud from previous interferometric observations \citep{Brick_jens, Brick_Katharine}. Using ALMA observations at 1\asec, \citet{rathborne15} estimated that this region contained 72~\msun within a radius of 0.04~pc, corresponding to a volume density of 3$\times$10$^{6}$~\cmq. If high-mass stars are presently forming anywhere in \brick, then this is the most likely location.

Our ALMA observations show that, while star formation is unambiguously underway in this region of the cloud, there are no obvious high-mass protostars residing within or in the vicinity of the \q{maser core}. The mass range of the detected cores is 0.6 -- 9.6~\msun \ with a median of 2~\msun, though this is subject to several caveats, particularly due to uncertain dust temperatures and missing flux (see section \hyperref[sec:continuum]{3.1}). We acknowledge that the upper limit for the masses of the central sources (\#1a \& b) is much larger (16 -- 18~\msun) due to the uncertainty in the dust temperature, which would put them in high-mass protostar territory. However, as discussed in section \hyperref[sec:gas_temp]{3.6.1}, their bright CH$_{3}$CN emission combined with their high densities means that it is likely that the dust and gas should be close to thermalised \citep{Clark_brick}, and their masses would therefore be on the lower end of the estimated range (2 -- 3~\msun). This is also consistent with the range of dynamical masses estimated in section \hyperref[sec:dynamical_mass]{3.6.2} (1.5 -- 5.6~\msun). 

In addition to the lack of high-mass cores, there are also no signatures of ongoing high-mass star formation. Maser surveys have only found water maser emission towards this region \citep[e.g.][]{Lu19}, but no class II methanol masers, which are typically indicative of high-mass star formation, have been detected. No HII regions have been detected towards the \q{maser core} in the radio continuum either \citep{Immer, Rod13, Mills15, Lu19B}. 

Despite the low-intermediate masses of the continuum sources in this region, the properties of the detected SiO outflows, namely their masses, energies, momenta, and mass outflow rates, are similar to those observed in intermediate and high-mass star-forming regions \citep[e.g.][and references therein]{Beuther02, Arce06, Zhang05, Bally_16, Beltran16}. These properties are also considered to be soft lower limits (see section \hyperref[sec:outflow_properties]{3.4}), and may be larger if the true SiO abundance is lower and if we are missing flux due to spatial filtering. The typical dynamical age is \ap 10$^{3}$ years, suggesting that the embedded protostars are young. This, coupled with the high outflow rates of \ap 10$^{-5}$ -- 10$^{-4}$~\msun~yr$^{-1}$, means that these sources may have the potential to become intermediate or high-mass stars in the future.

Of particular interest in this context are the two central sources in this field. The cores are situated in a larger-scale clump of dense material (see Figure \hyperref[fig:full_continuum]{2}). Assuming a dust temperature of 22~K and subtracting the flux contribution from the embedded sources, we estimate that this envelope has a mass of \ap 35~\msun \ within 5000~AU. If the embedded protostars were able to efficiently accrete from this material and the larger scale reservoir, they could potentially grow to become high-mass stars. The lower limit to the total mass outflow rate of these sources is \ap 2$\times$10$^{-4}$~\msun~yr$^{-1}$, which is consistent with outflow rates observed in high-mass star-forming regions. Assuming that this is a lower limit to the mass infall rate, and that the accretion rate onto the protostar(s) is some fraction of the infall rate, then it would take 10$^{4-5}$ yrs to form a \ap 10~\msun \ star, and potentially less if the accretion rate is variable and increases as the protostars grow in mass \citep[e.g][]{Zhang05, Zhang15, Li_20}. Given the densities presented in Table \hyperref[tab:cores]{3}, the typical free-fall time of the detected sources is a few thousand years. This means that in the aforementioned scenario, the central sources would have to accrete over many free-fall times.

With the data presented in this paper, we are not able to resolve the central sources well, and it is not possible to determine the contribution that either of the sources is making to the outflowing material. Follow-up molecular line observations at higher angular resolution are required to better constrain the infall and accretion rates onto the central protostars, and ultimately determine whether this could represent the early stages of a high-mass binary in \brick. 

\section{Conclusions}
\label{sec:conclusion}
We have presented high-resolution (0.13\asec, 1000~AU) ALMA Band 6 (1.3~mm, 230~GHz) observations towards the \q{maser core} in the extreme Galactic centre cloud \brick. The main results are summarised as follows:

\begin{itemize}[leftmargin=0.25in]
  \item The \q{maser core} region fragments significantly on 1000~AU scales, revealing a small cluster of at least 18 compact sources that are detected in the dust continuum. The median mass of the cores is 2~\msun, with typical radii of \ap 1000~AU and densities of 10$^{7-8}$~\cmq.

\medskip

  \item We detect at least 9 bi-polar outflows via SiO (5-4) emission that are associated with the observed dust continuum sources. We also find potential evidence for outflows and outflow cavities traced by $^{13}$CO (2-1) emission. This constitutes unambiguous evidence of active star formation in \brick.

\medskip

  \item The central source of the \q{maser core} dominates the continuum flux on small-scales, and is revealed to be a protostellar binary system (projected separation \ap 1000~AU), with possible signs of higher multiplicity. This system is driving at least one collimated outflow, with several other multi-directional lobes, indicating that there are likely multiple sources accreting.

\medskip

  \item Despite the high densities towards this region (\textgreater \ 10$^{6}$~\cmq) and the large reservoir of dense gas in \brick \ as a whole, we find no evidence of high-mass protostars. However, the observed SiO outflow properties are consistent with those observed towards intermediate and high-mass protostars, and so some of the detected cores may potentially grow to become high-mass stars in the future. The central protostellar binary is a promising candidate for a future high-mass stellar binary, as it is embedded in a dense reservoir of material. Direct measurements of infall/accretion rates are necessary to determine whether they could potentially become high-mass stars in the future.

\medskip

  \item The masses and distribution of the detected continuum sources are found to be generally consistent with thermal Jeans fragmentation. This suggests that the large-scale turbulence may not play a significant role in shaping the cloud structure on protostellar scales, and that the mechanisms governing the fragmentation of protostellar-scale structure in this extreme Galactic environment are similar to those in the nearby star-forming regions at the individual core scale.
  
\end{itemize}

\section*{Acknowledgements}

This paper makes use of the following ALMA data: ADS/JAO.ALMA\#2016.1.00949.S. ALMA is a partnership of ESO (representing its member states), NSF (USA) and NINS (Japan), together with NRC (Canada), MOST and ASIAA (Taiwan), and KASI (Republic of Korea), in cooperation with the Republic of Chile. The Joint ALMA Observatory is operated by ESO, AUI/NRAO and NAOJ. DLW and CB gratefully acknowledge support from the National Science Foundation under Award No. 1816715. A.G. acknowledges support from the National Science Foundation under grant No. 2008101. JMDK gratefully acknowledges funding from the German Research Foundation (DFG) in the form of an Emmy Noether Research Group (grant number KR4801/1-1) and from the European Research Council (ERC) under the European Union's Horizon 2020 research and innovation programme via the ERC Starting Grant MUSTANG (grant agreement number 714907). ATB would like to acknowledge funding from the European Research Council (ERC) under the European Union’s Horizon 2020 research and innovation programme (grant agreement No.726384/Empire). LCH was supported by the National Key R\&D Program of China (2016YFA0400702) and the  National Science Foundation of China (11721303, 11991052). HB acknowledges funding from the European Research Council under the Horizon 2020 Framework Program via the ERC Consolidator Grant CSF-648505. HB also acknowledges support from the Deutsche Forschungsgemeinschaft via SFB 881, “The Milky Way System” (subproject B1). XL acknowledges the financial support by JSPS KAKENHI grants No. 18K13589 \& 20K14528. The authors thank Antoine Gusdorf for their helpful discussion concerning the SiO abundances in this region. The authors thank the anonymous referee for their comments, which improved the quality and clarity to this manuscript.

\bigskip
\noindent
{\bf{Software}}: This research primarily made use of the following software packages: \textbf{CASA} \citep{casa}, \textbf{astropy}, a community-developed core Python package for Astronomy \citep{astropy:2013, astropy:2018}, \textbf{astrodendro}, a Python package to compute dendrograms of astronomical data (\url{http://www.dendrograms.org/}), \textbf{APLpy}, an open-source plotting package for Python \citep{aplpy}, \textbf{spectral-cube} (\url{https://spectral-cube.readthedocs.io/en/latest/}), \textbf{radio-beam} (\url{https://radio-beam.readthedocs.io/en/latest/}), \textbf{statcont} (\url{https://hera.ph1.uni-koeln.de/~sanchez/statcont}), and \textbf{XCLASS} (\url{https://xclass.astro.uni-koeln.de/}).}

\bigskip
\noindent
{\bf{Data availability}}: The data products used to conduct the research presented in this paper are made publicly available at the following Harvard Dataverse repository:\\ \url{https://doi.org/10.7910/DVN/FXORIW}. 

\noindent
Uncompressed versions of the figures included in the paper are also made available in the same Dataverse at:\\ \url{https://doi.org/10.7910/DVN/O6GN1T}.

\bibliography{brick_paper_bib}



\begin{figure*}
\begin{subfigure}[b]{0.49\textwidth} 
    \includegraphics[width=\linewidth]{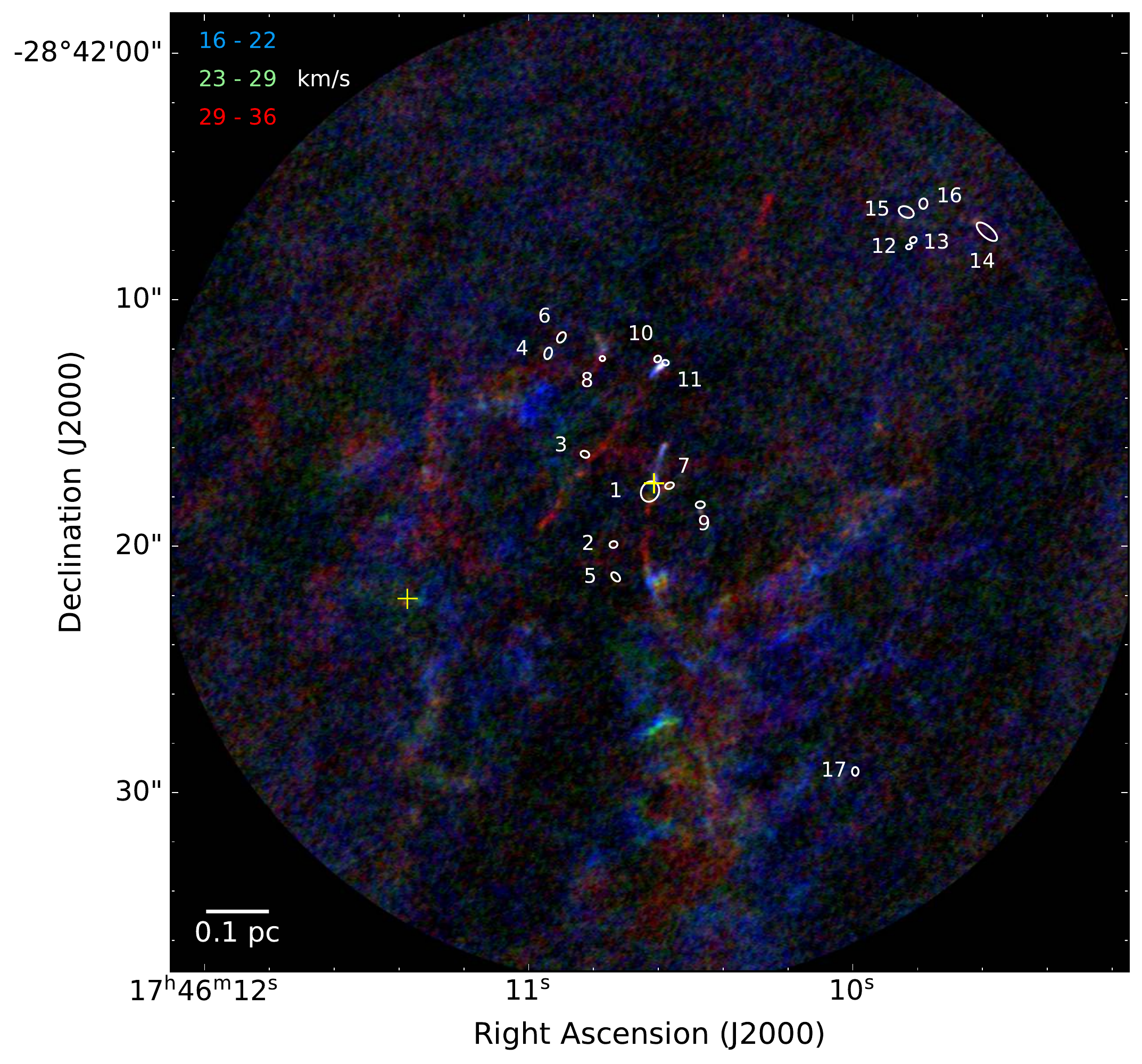}
    \caption{} 
\end{subfigure}%
\hfill
\begin{subfigure}[b]{0.49\textwidth}
    \centering
    \includegraphics[width=\linewidth]{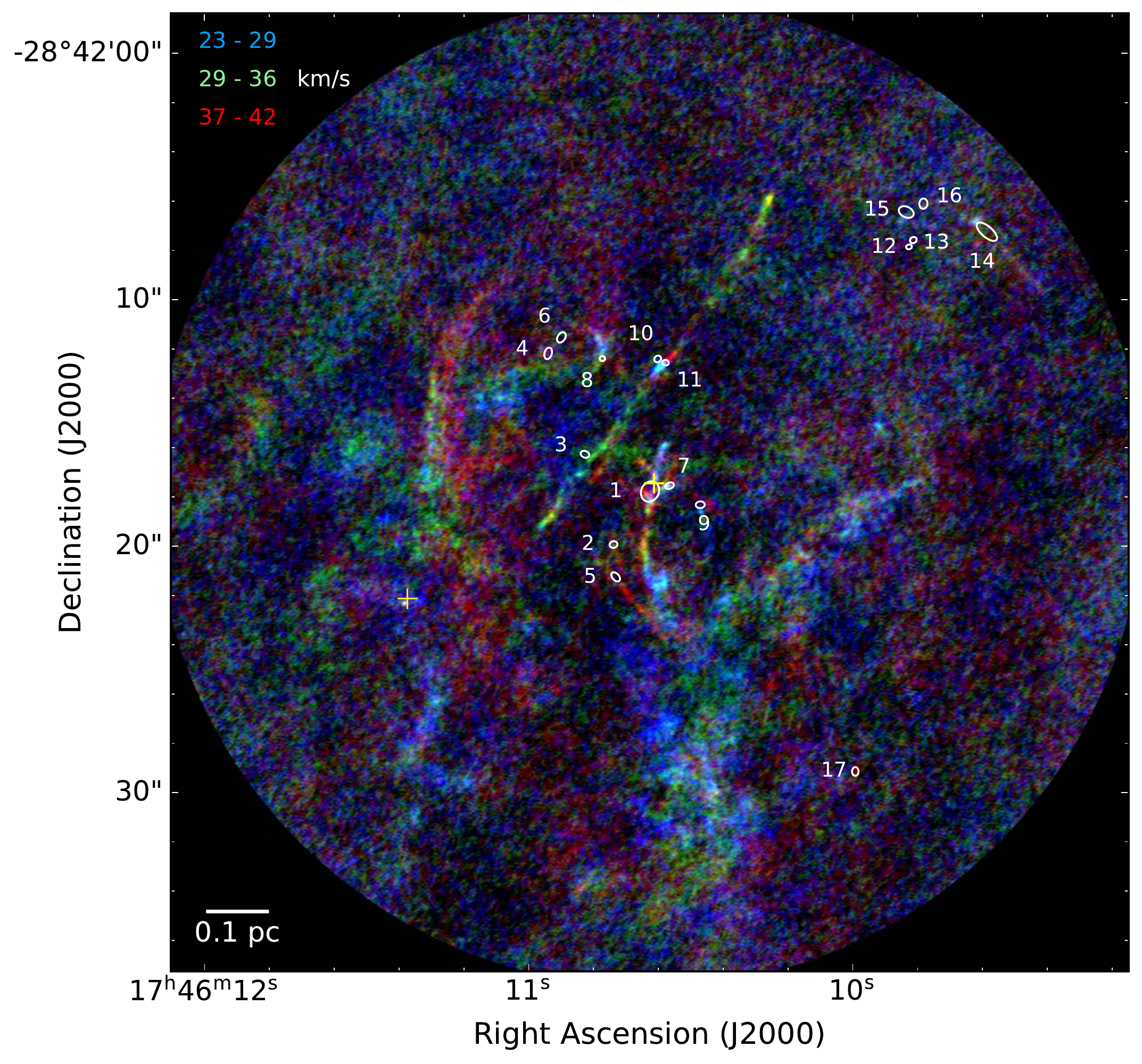}
    \caption{} 
    \end{subfigure}
    
\begin{subfigure}[c]{0.49\textwidth} 
    \includegraphics[width=\linewidth]{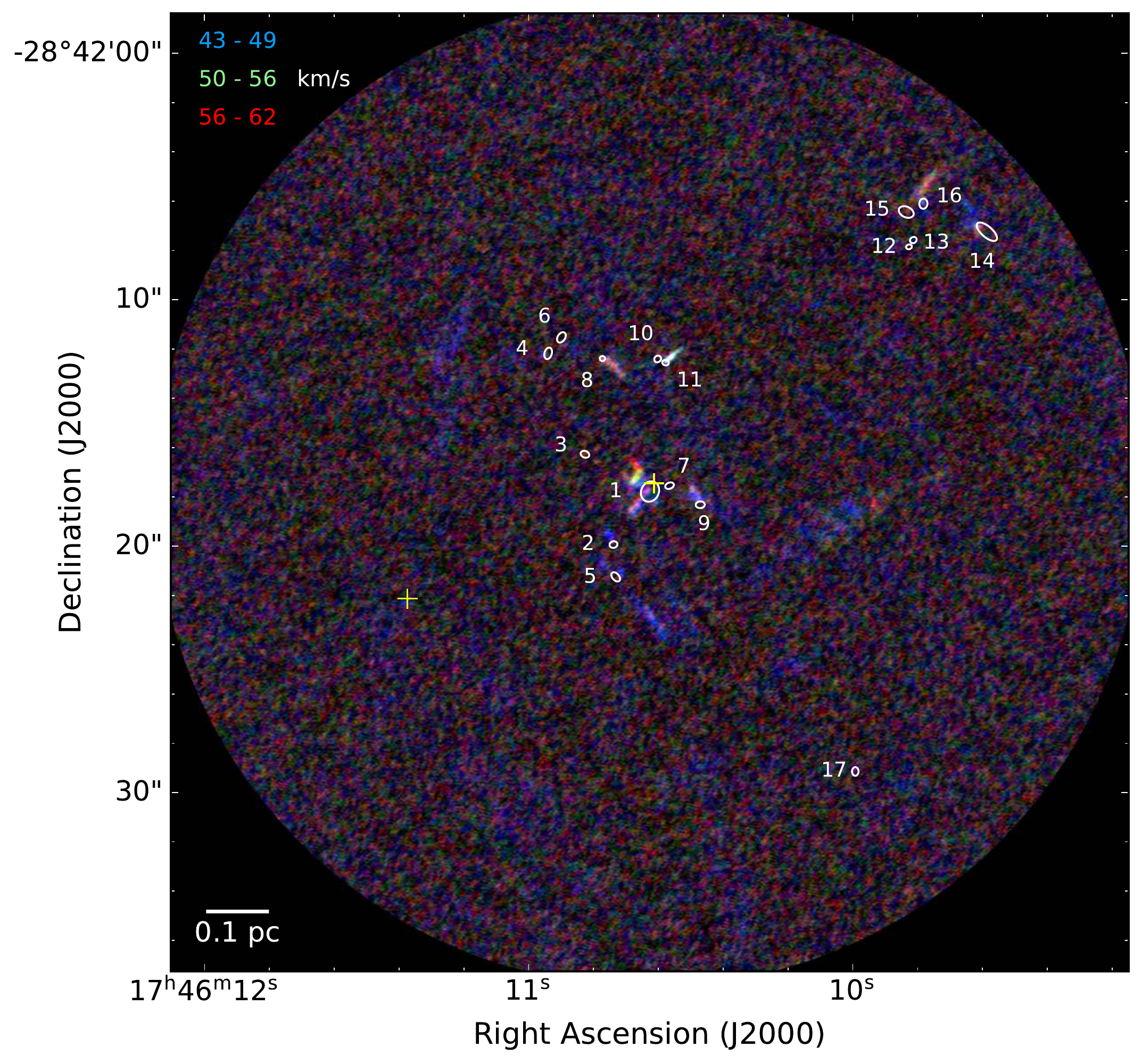}
    \caption{} 
\end{subfigure}%
\hfill
\begin{subfigure}[d]{0.49\textwidth}
    \centering
    \includegraphics[width=\linewidth]{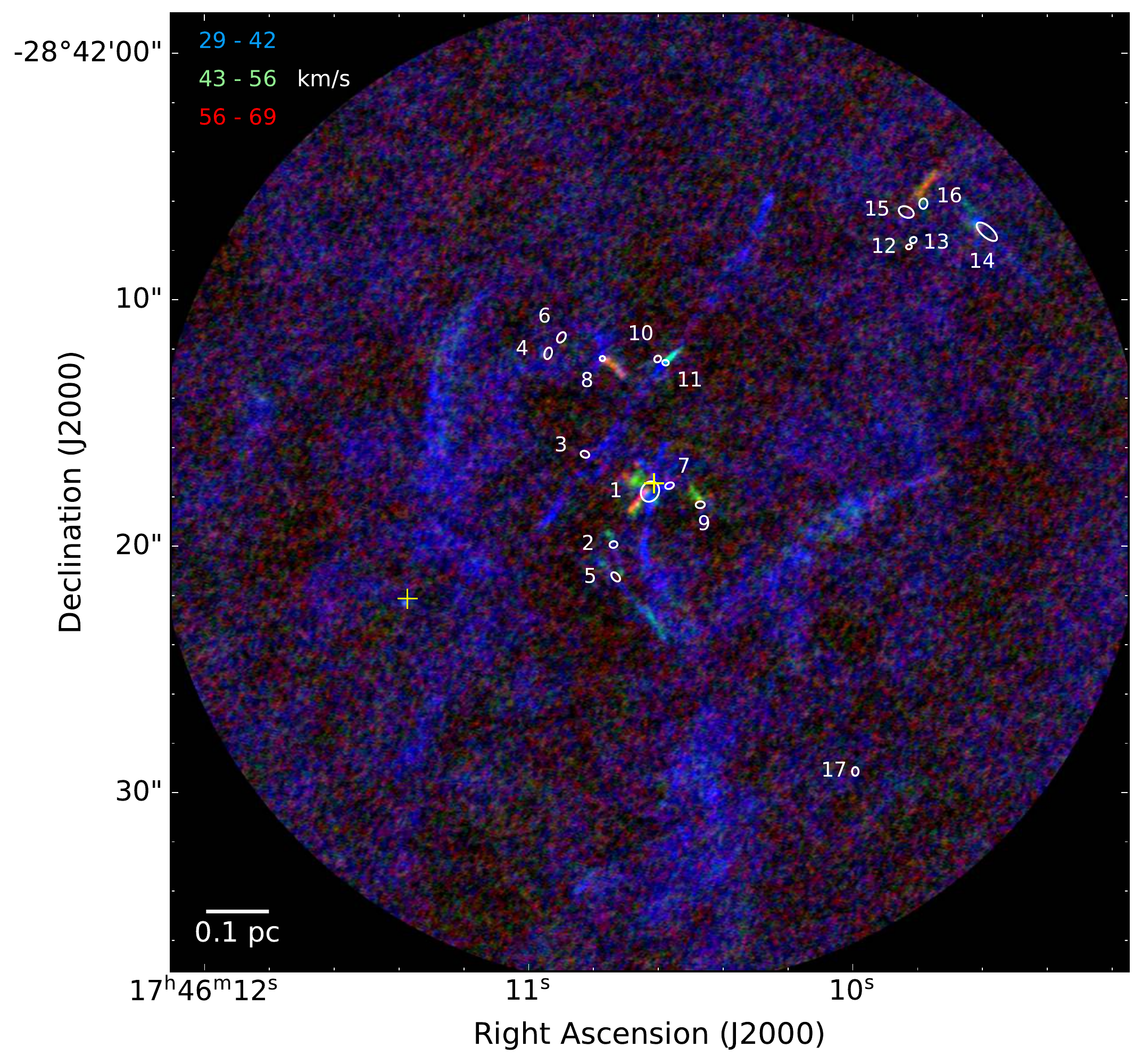}
    \caption{} 
    \end{subfigure}
    \caption{Three-colour figures showing the SiO (5-4) emission at different velocities. \textbf{(a)}: 16 -- 22 (blue), 23 -- 29 (green), and 29 -- 36 (red) \kms, \textbf{(b)}: 23 -- 39 (blue), 29 -- 36 (green), and 37 -- 42 (red) \kms, \textbf{(c)}: 43 -- 49 (blue), 50 -- 56 (green), and 56 -- 62 (red) \kms, \textbf{(d)}: 29 -- 42 (blue), 43 -- 56 (green), and 56 -- 69 (red) \kms. Continuum sources are highlighted by white ellipses, the extent of which corresponds to the structures determined using dendrograms. Each continuum source is also numbered. The yellow crosses show the position of water masers from \citet{Lu19}.} 
\label{fig:SiO_appendix1}
\end{figure*}

\begin{figure*}
\begin{subfigure}[b]{0.49\textwidth} 
    \includegraphics[width=\linewidth]{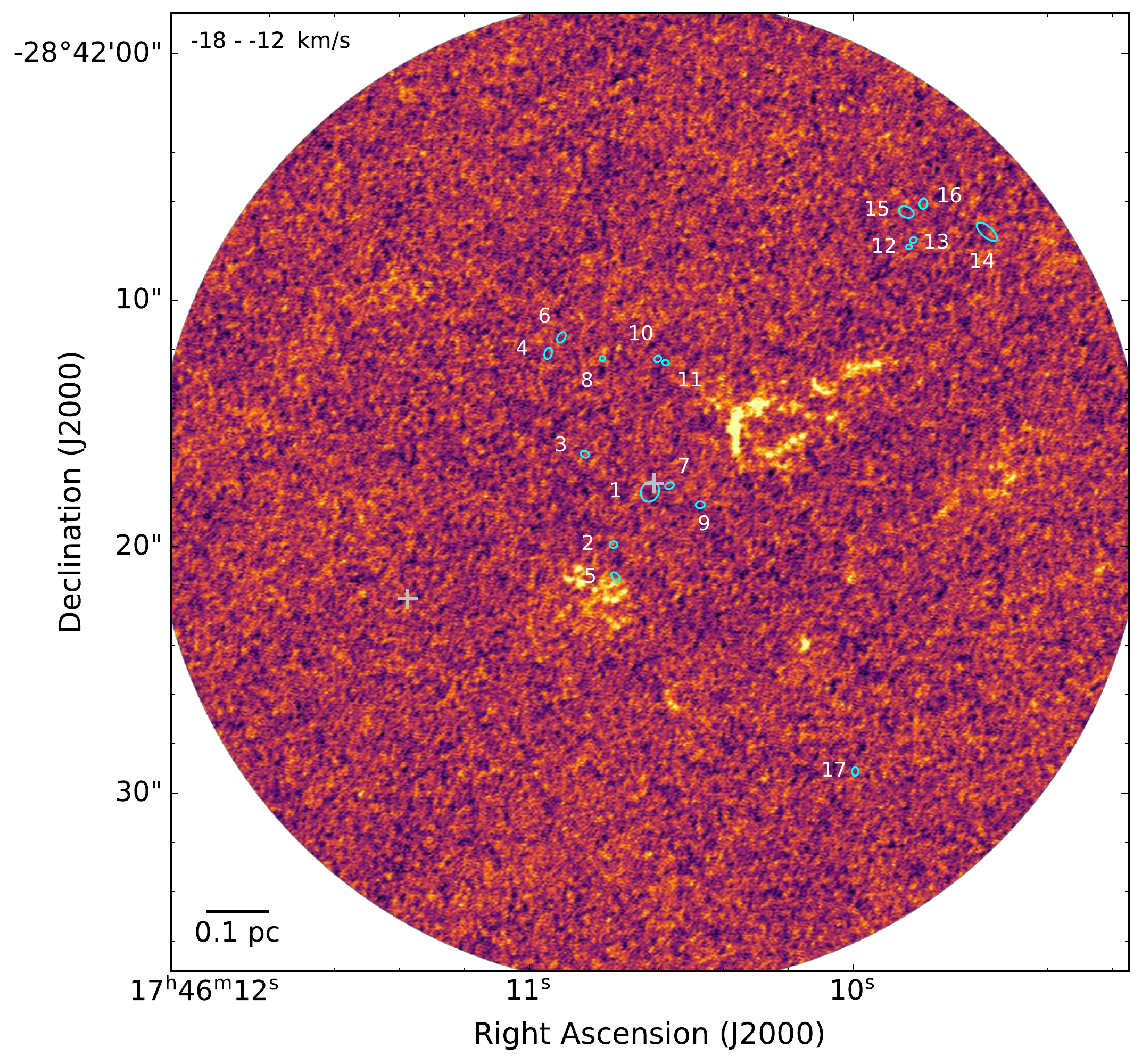}
    \caption{} 
\end{subfigure}%
\hfill
\begin{subfigure}[b]{0.49\textwidth}
    \centering
    \includegraphics[width=\linewidth]{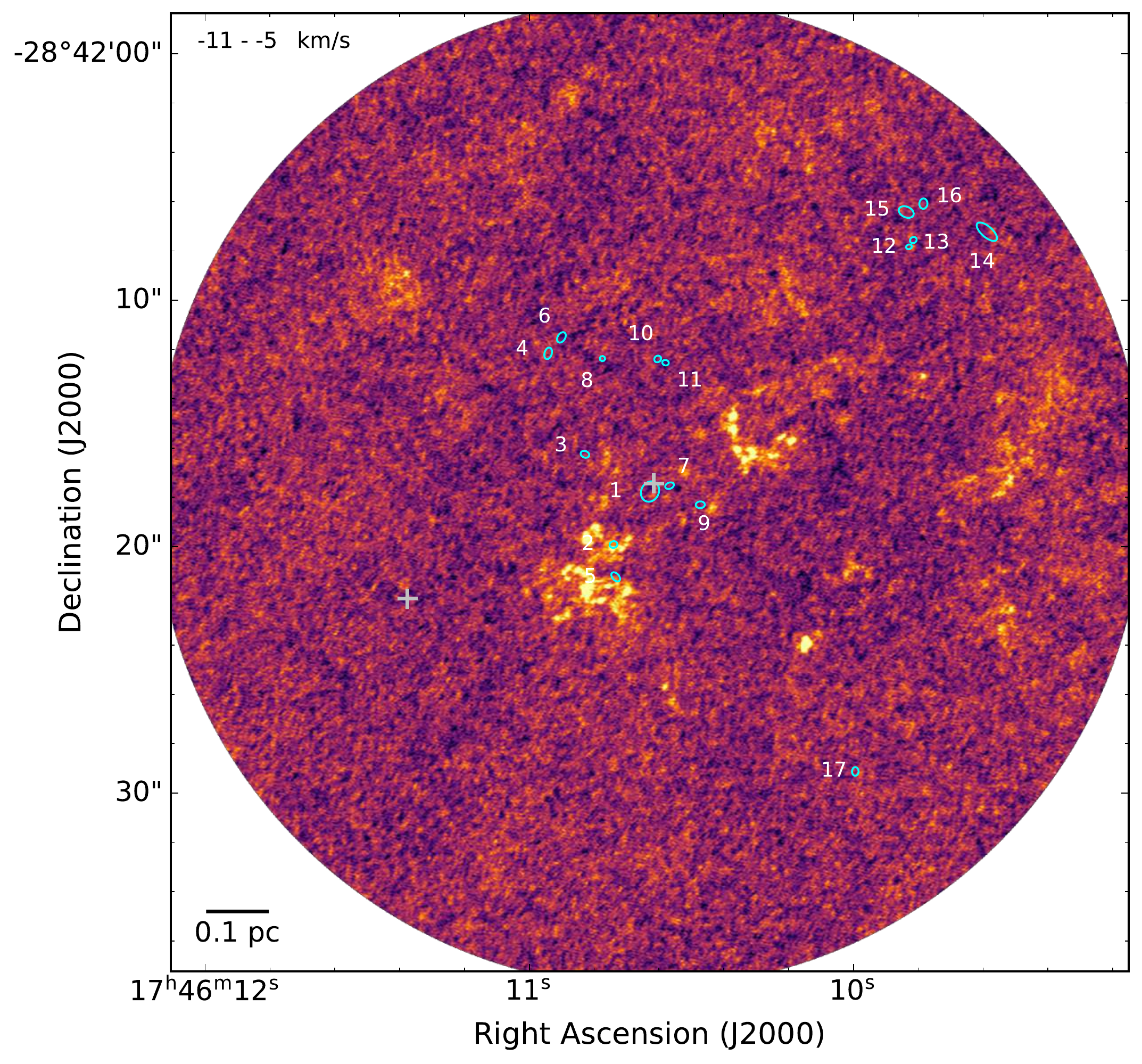}
    \caption{} 
    \end{subfigure}
    
\begin{subfigure}[c]{0.49\textwidth} 
    \includegraphics[width=\linewidth]{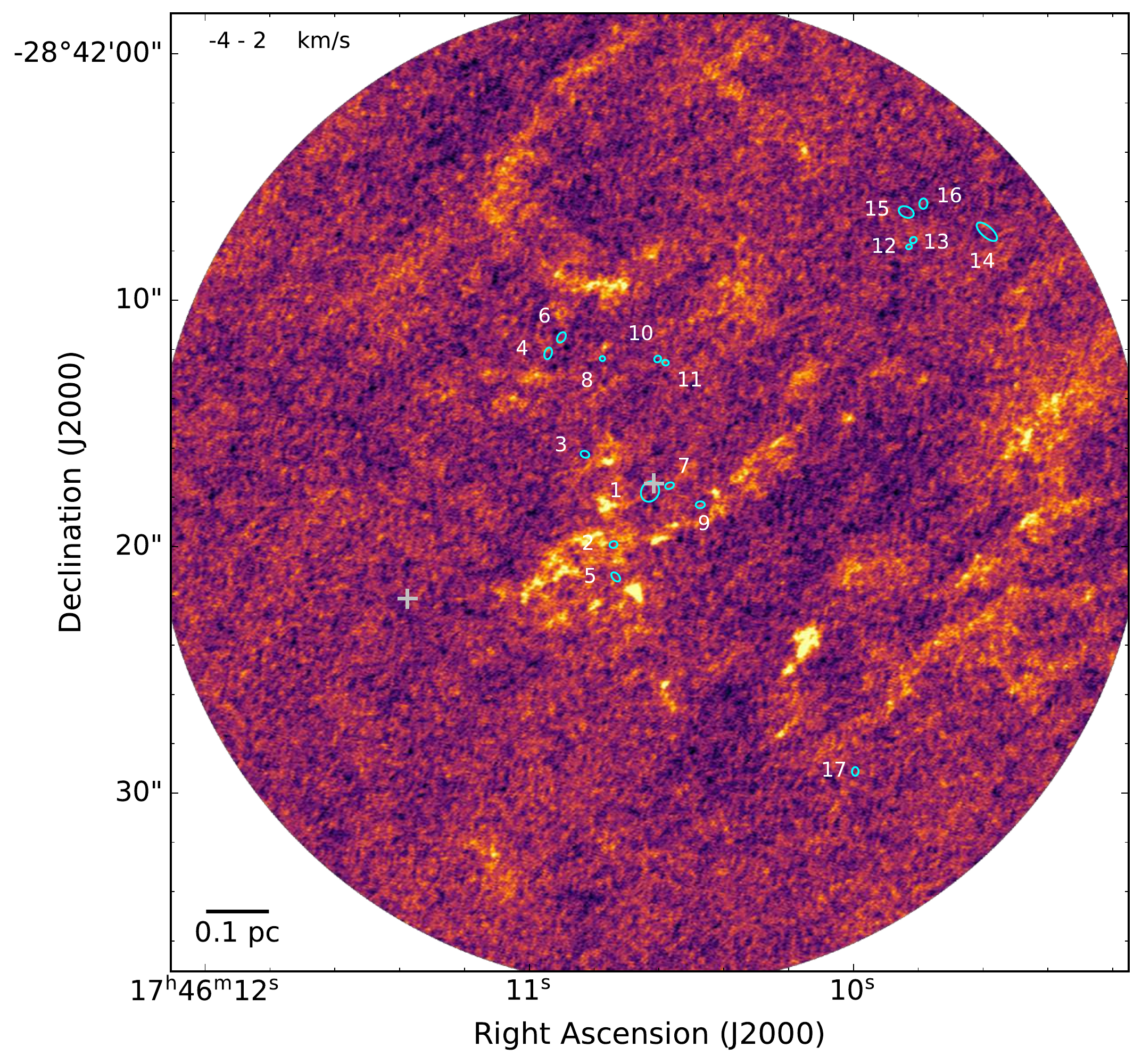}
    \caption{} 
\end{subfigure}%
\hfill
\begin{subfigure}[d]{0.49\textwidth}
    \centering
    \includegraphics[width=\linewidth]{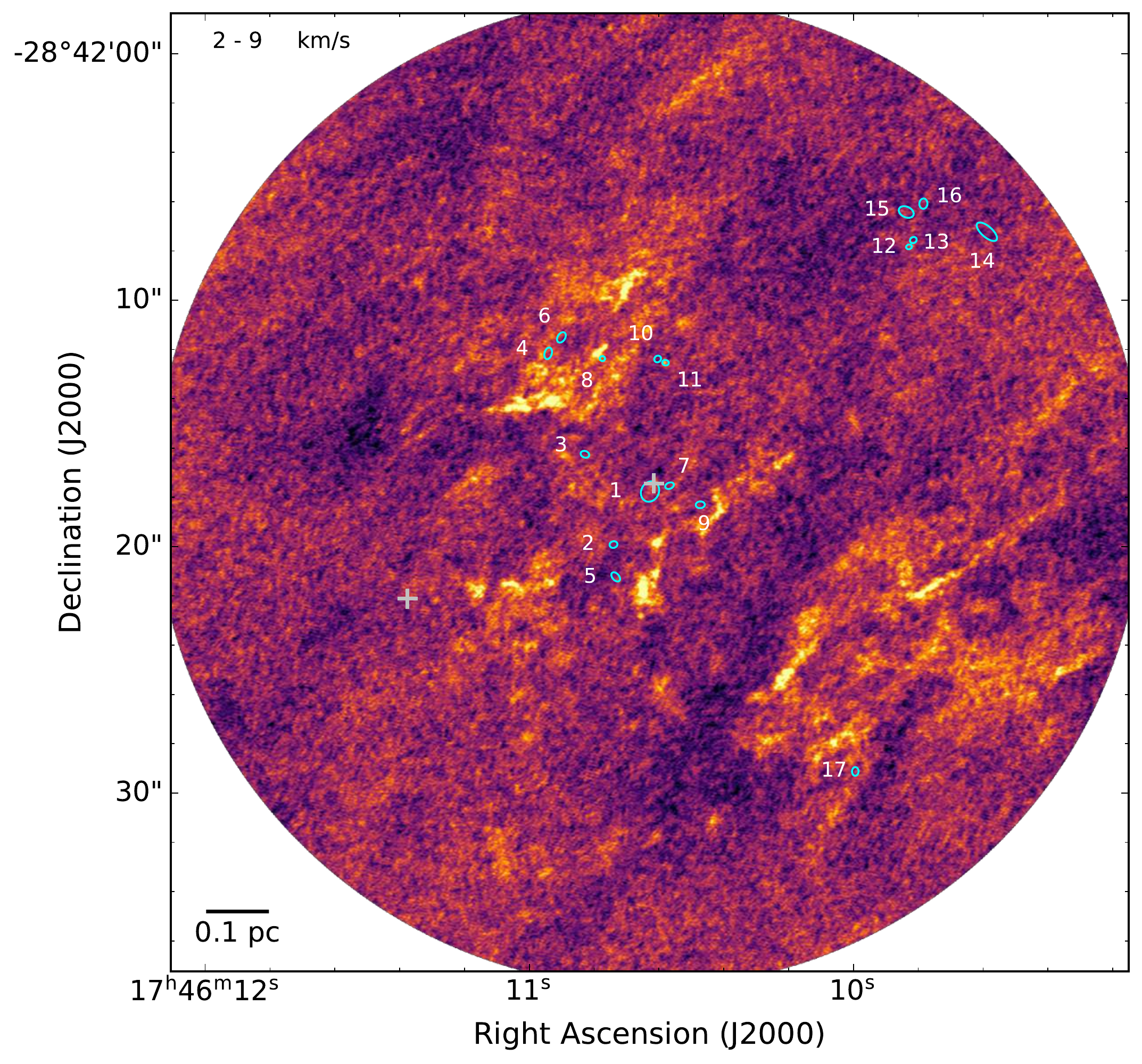}
    \caption{} 
    \end{subfigure}
    \caption{Figures showing the integrated SiO (5-4) emission in different velocity ranges. \textbf{(a)}: -18 -- -12 \kms, \textbf{(b)}: -11 -- -5 \kms, \textbf{(c)}: -4 -- 2 \kms, \textbf{(d)}: 2 -- 9 \kms. Continuum sources are highlighted by cyan ellipses, the extent of which corresponds to the structures determined using dendrograms. Each continuum source is also numbered. The grey crosses show the position of water masers from \citet{Lu19}.} 
\label{fig:SiO_appendix2}
\end{figure*}

\begin{figure*}
\begin{subfigure}[b]{0.49\textwidth} 
    \includegraphics[width=\linewidth]{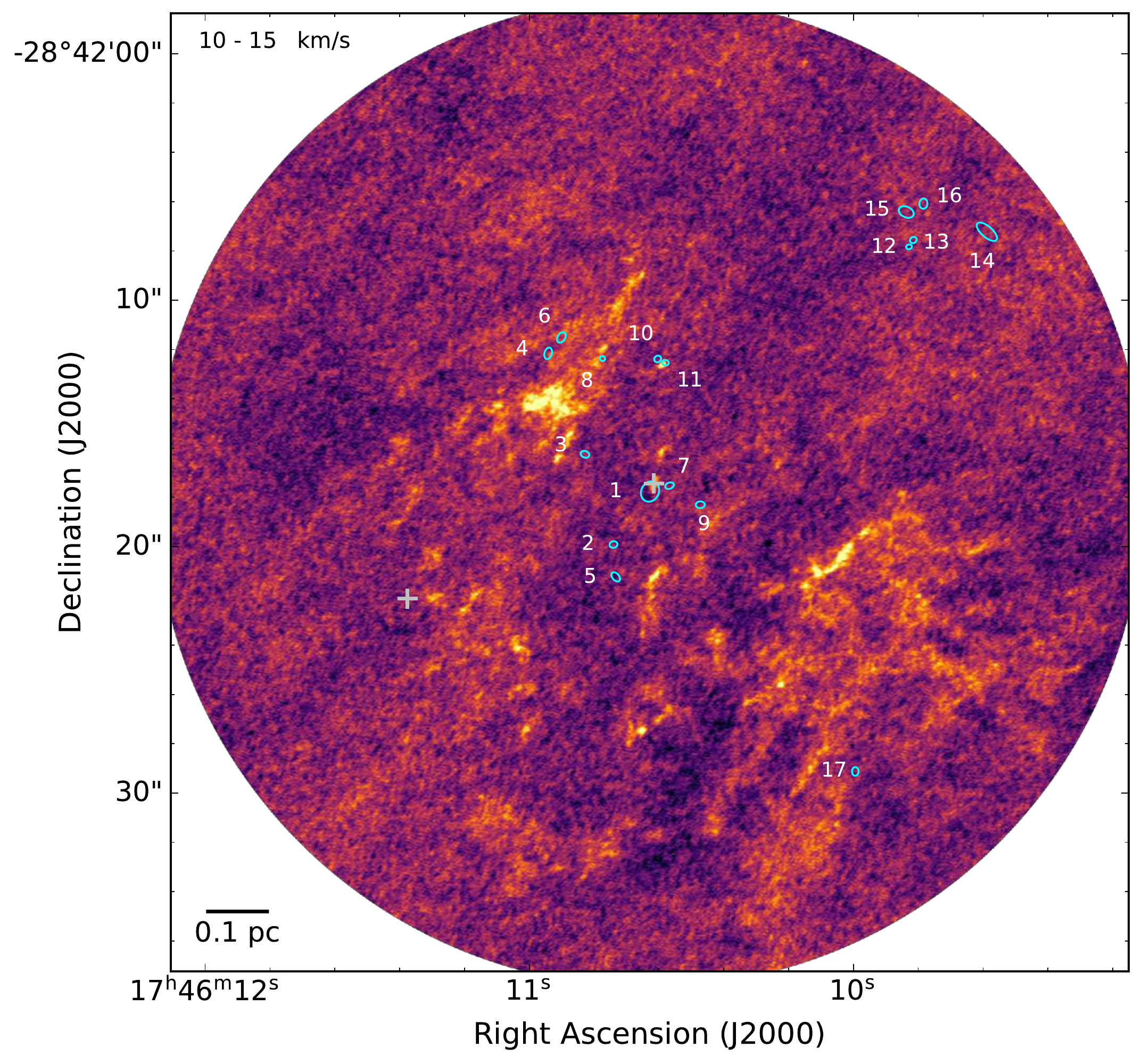}
    \caption{} 
\end{subfigure}%
\hfill
\begin{subfigure}[b]{0.49\textwidth}
    \centering
    \includegraphics[width=\linewidth]{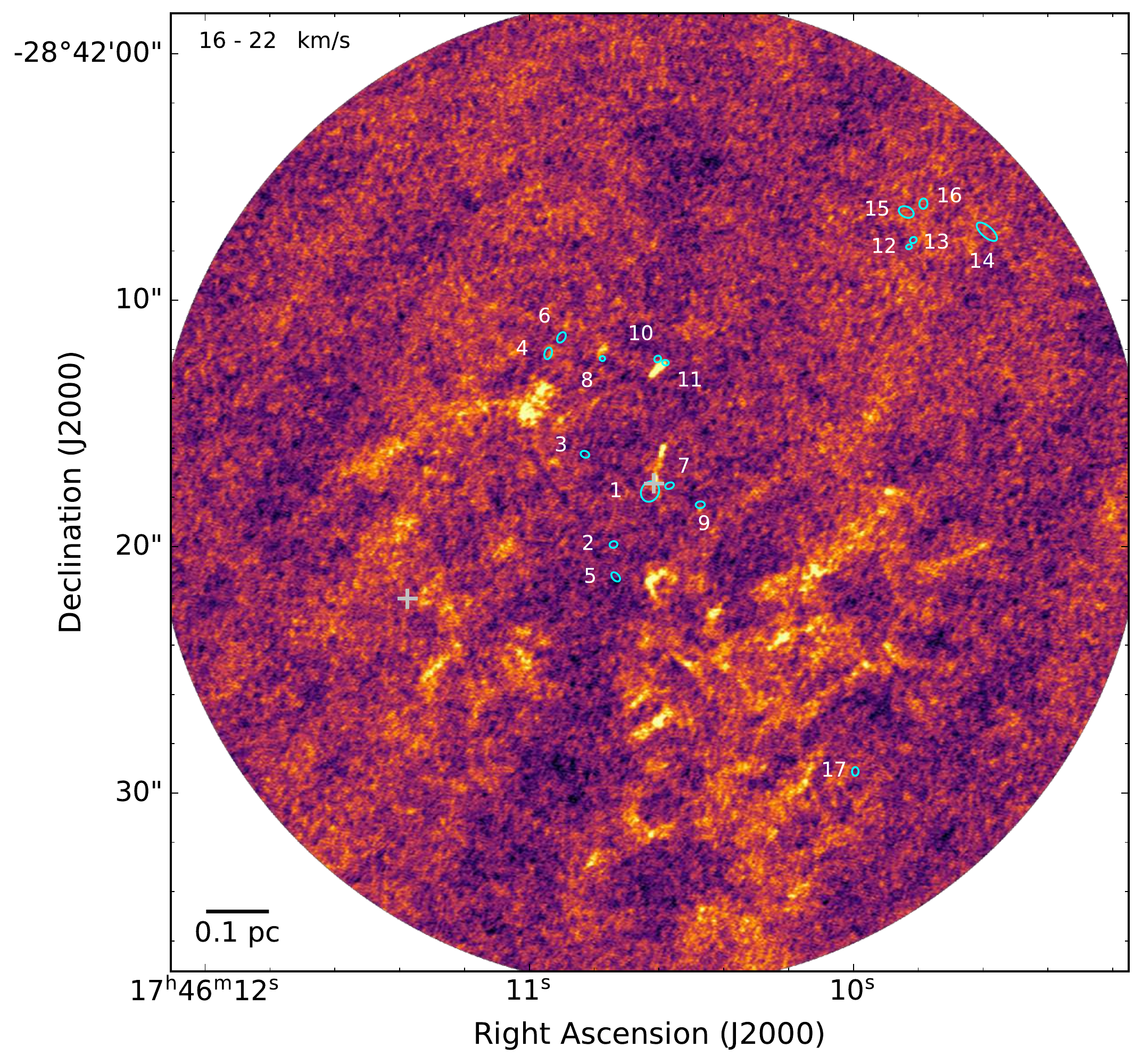}
    \caption{} 
    \end{subfigure}
    
\begin{subfigure}[c]{0.49\textwidth} 
    \includegraphics[width=\linewidth]{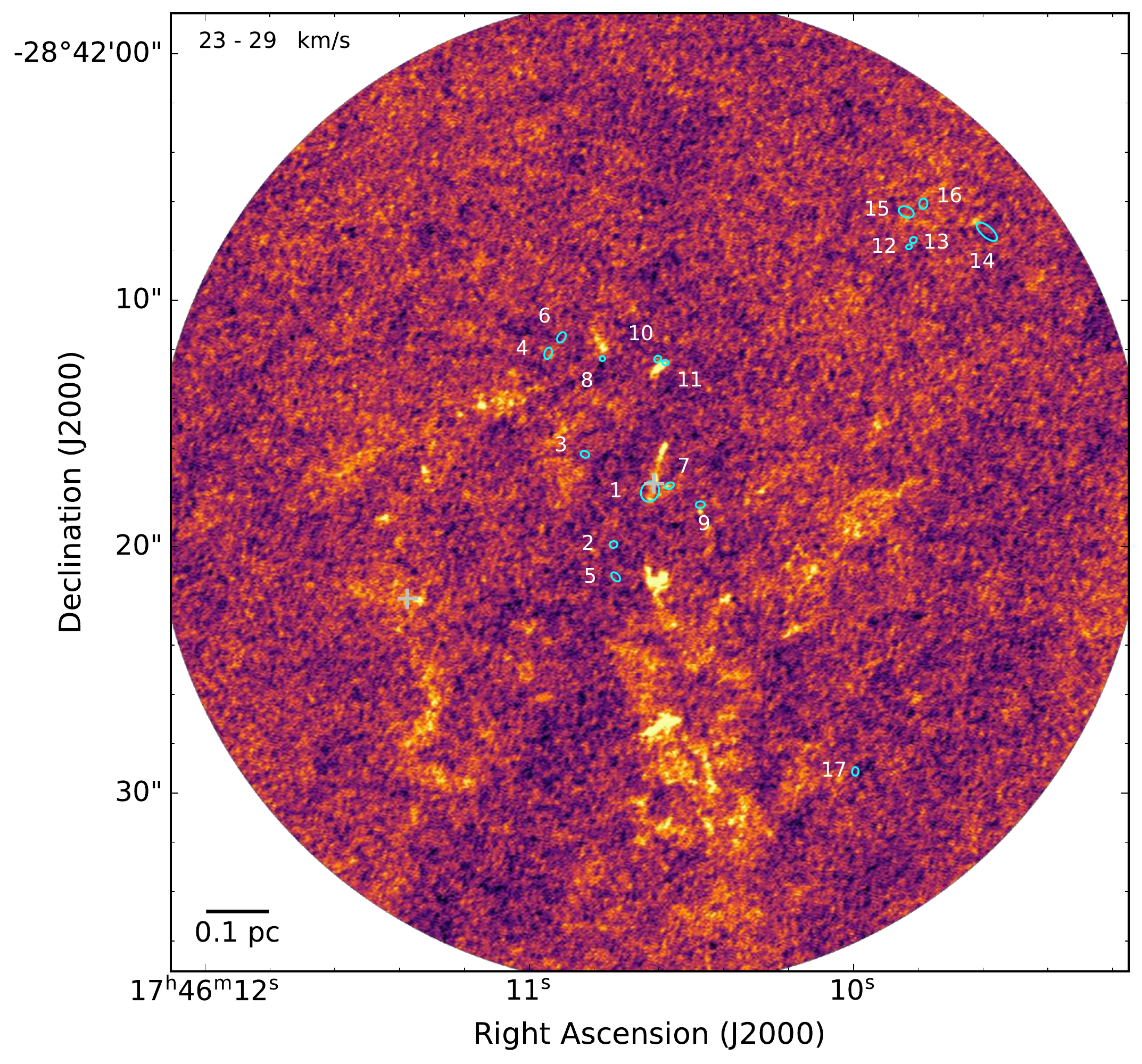}
    \caption{} 
\end{subfigure}%
\hfill
\begin{subfigure}[d]{0.49\textwidth}
    \centering
    \includegraphics[width=\linewidth]{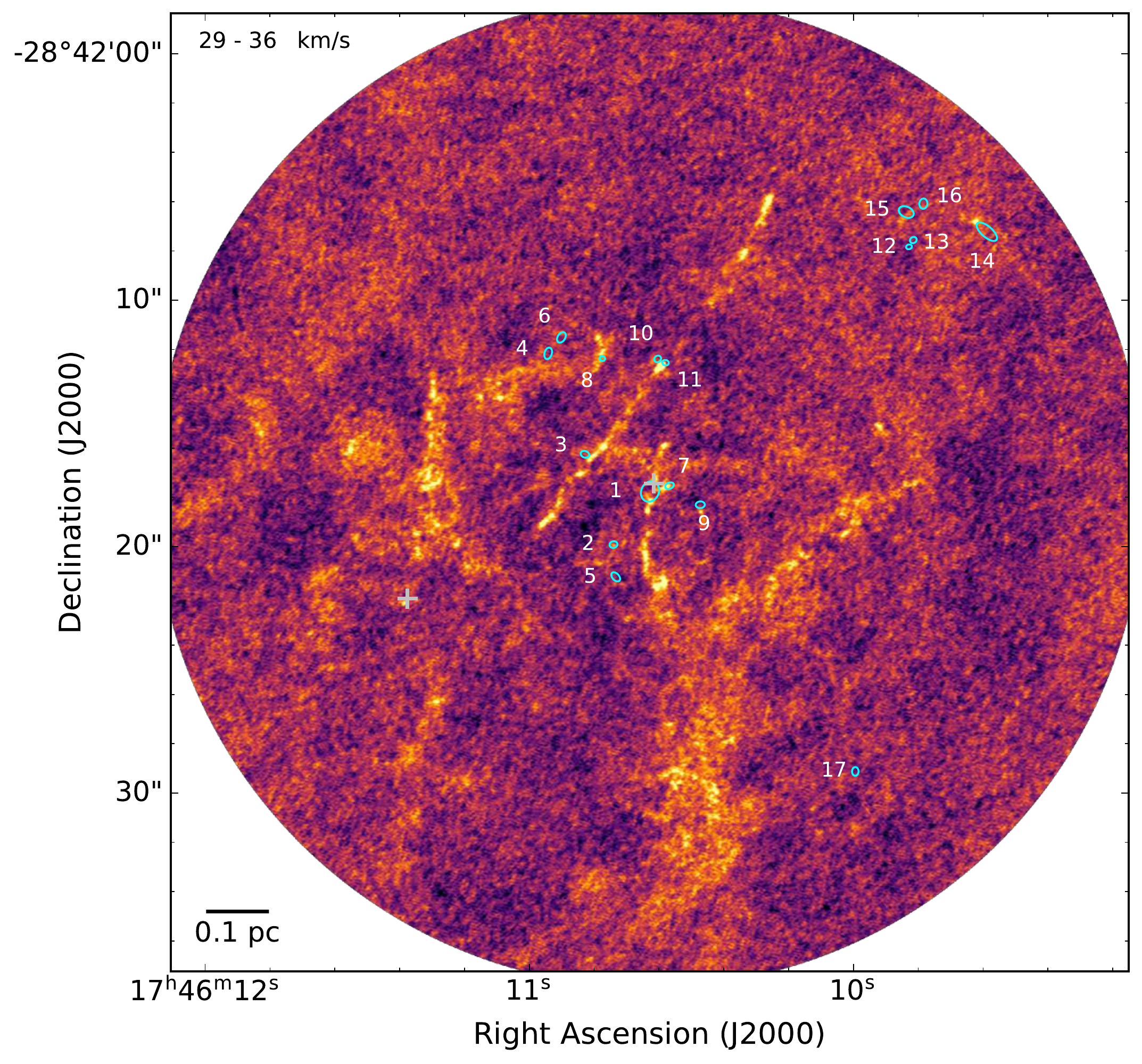}
    \caption{} 
    \end{subfigure}
    \caption{As in Figure 10. \textbf{(a)}: 10 -- 15 \kms, \textbf{(b)}: 16 -- 22 \kms, \textbf{(c)}: 23 -- 29 \kms, \textbf{(d)}: 29 -- 36 \kms.} 
\label{fig:SiO_appendix3}
\end{figure*}

\begin{figure*}
\begin{subfigure}[b]{0.49\textwidth} 
    \includegraphics[width=\linewidth]{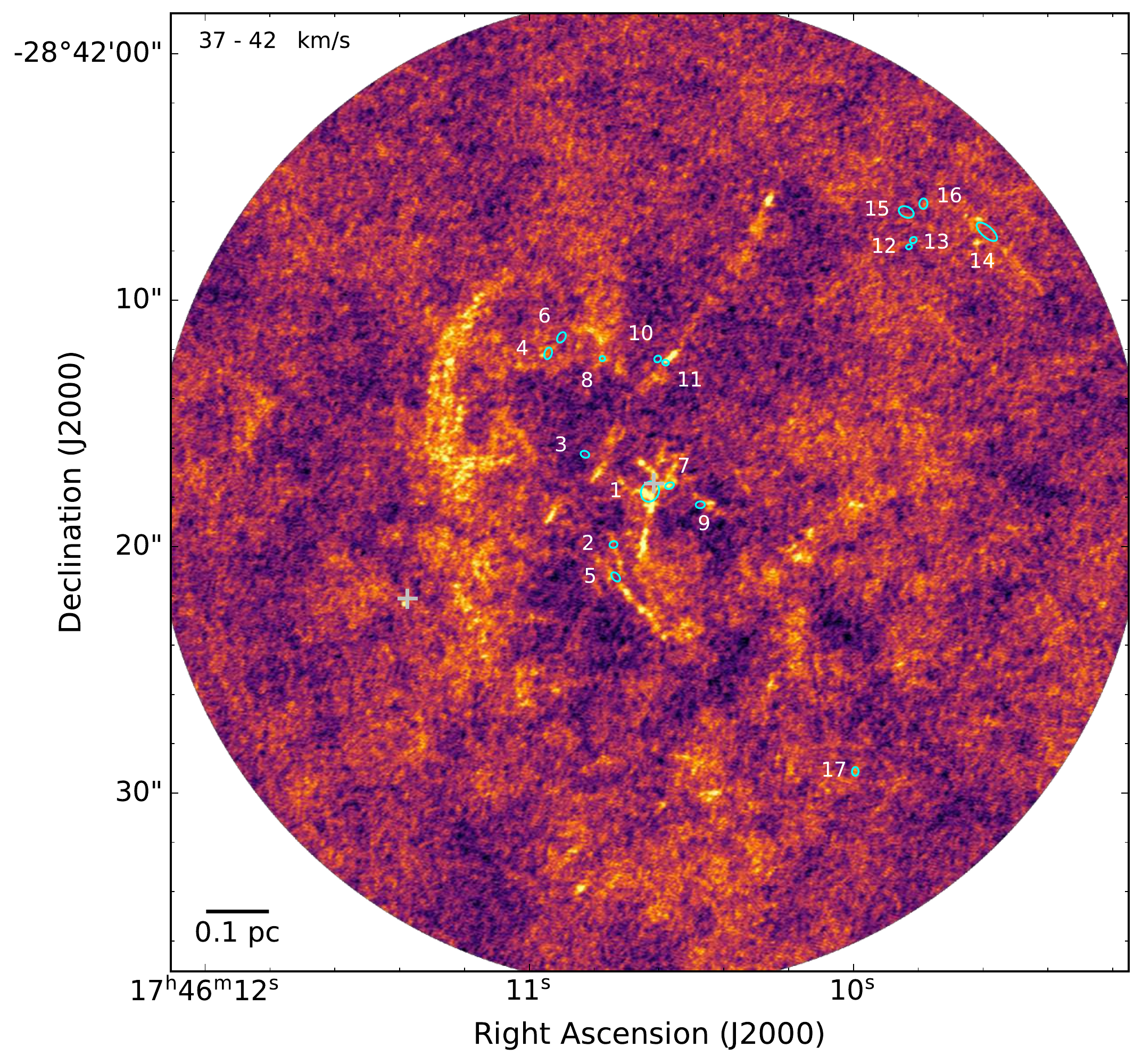}
    \caption{} 
\end{subfigure}%
\hfill
\begin{subfigure}[b]{0.49\textwidth}
    \centering
    \includegraphics[width=\linewidth]{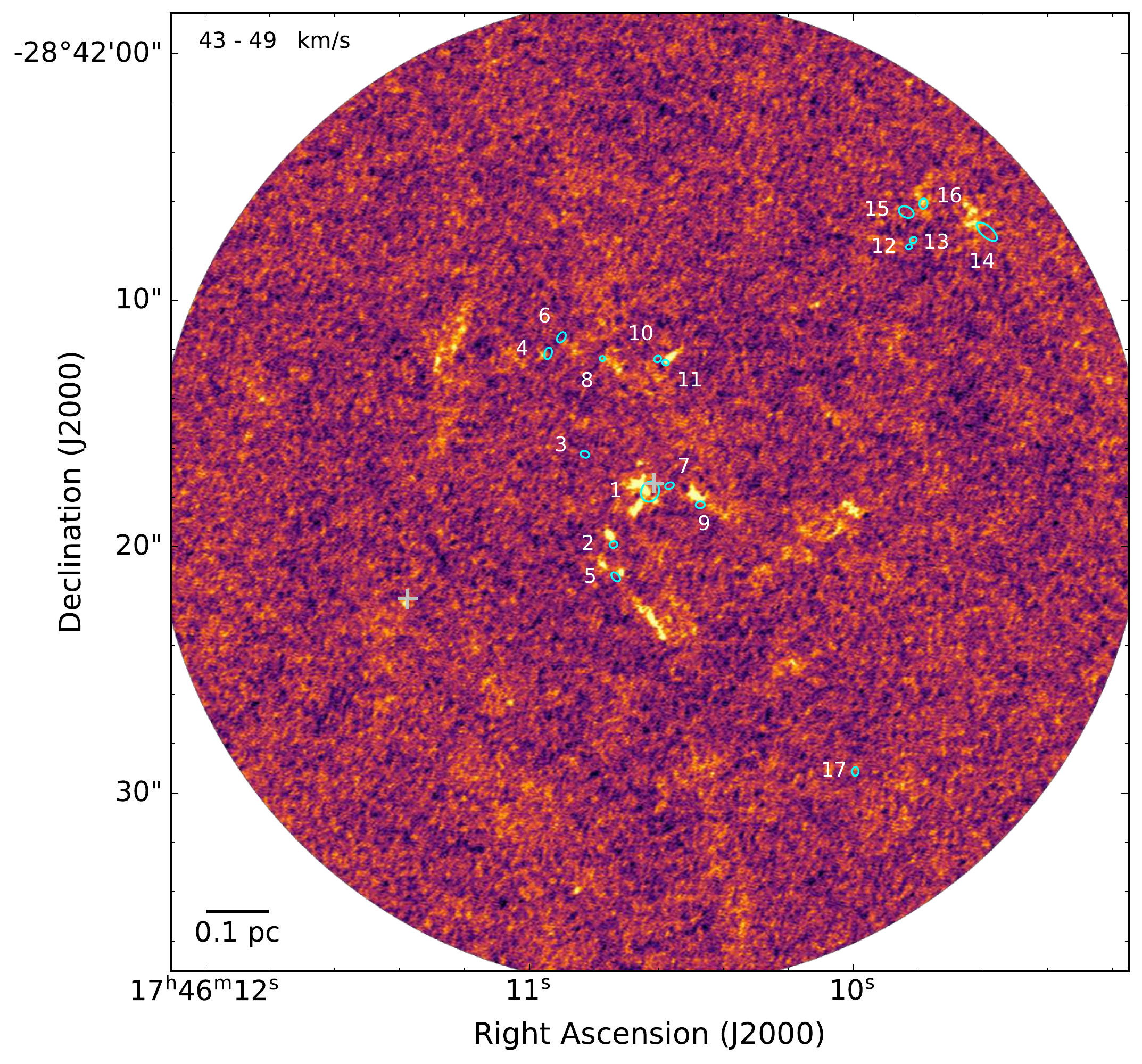}
    \caption{} 
    \end{subfigure}
    
\begin{subfigure}[c]{0.49\textwidth} 
    \includegraphics[width=\linewidth]{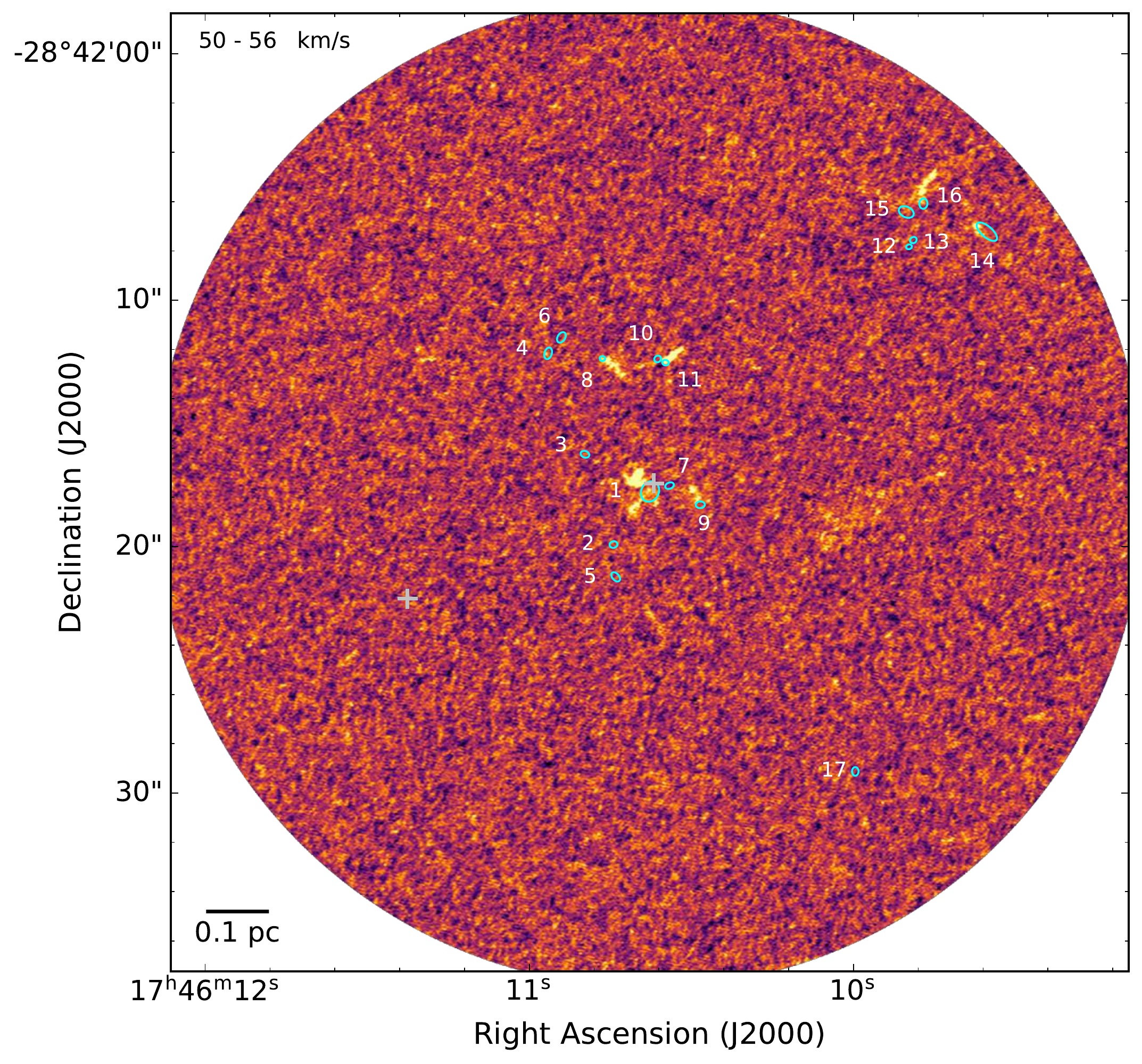}
    \caption{} 
\end{subfigure}%
\hfill
\begin{subfigure}[d]{0.49\textwidth}
    \centering
    \includegraphics[width=\linewidth]{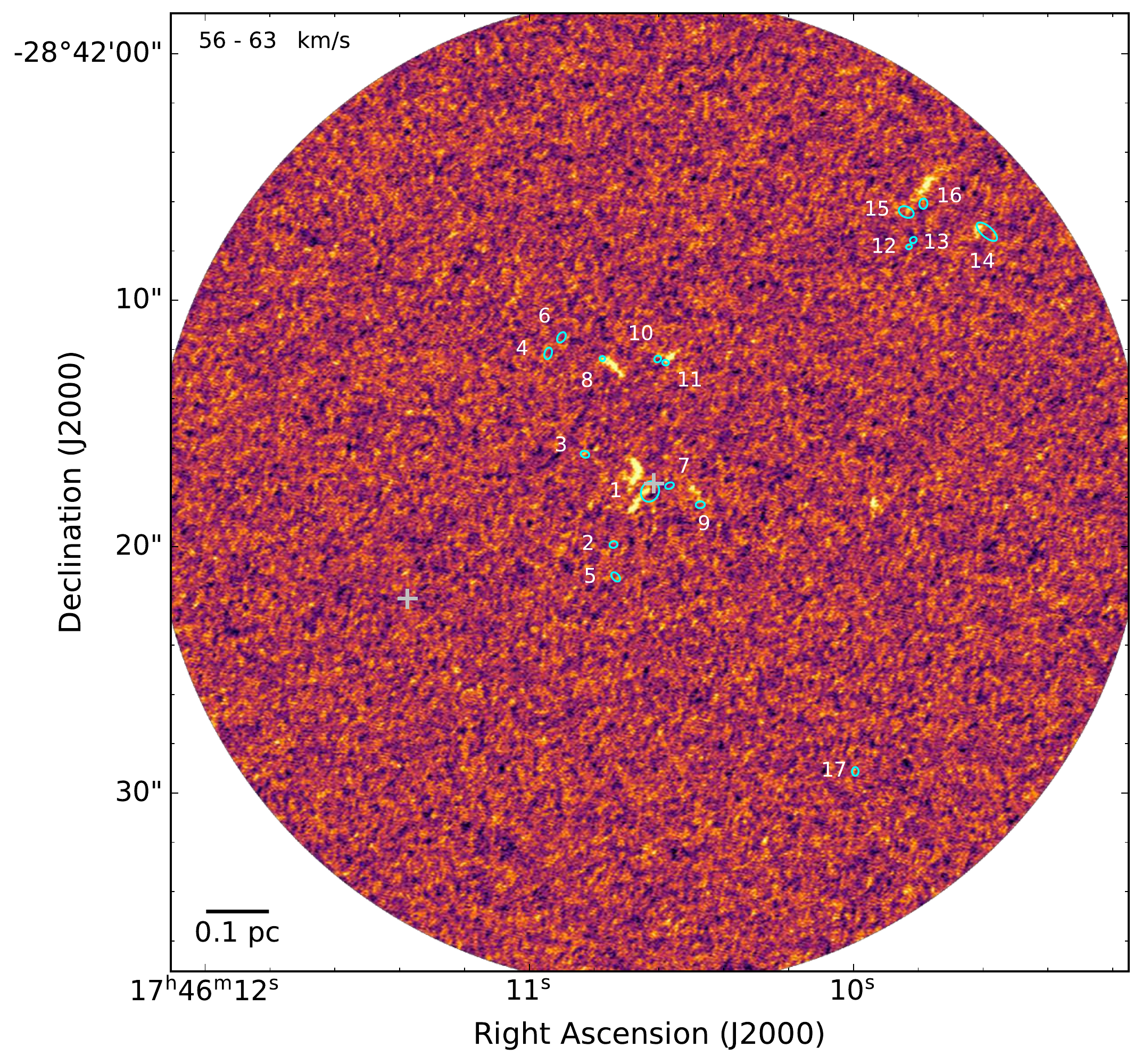}
    \caption{} 
    \end{subfigure}
    \caption{As in Figure 10. \textbf{(a)}: 37 -- 42 \kms, \textbf{(b)}: 43 -- 49 \kms, \textbf{(c)}: 50 -- 56 \kms, \textbf{(d)}: 56 -- 62 \kms.} 
\label{fig:SiO_appendix4}
\end{figure*}

\begin{figure*}
\begin{subfigure}[b]{0.49\textwidth} 
    \includegraphics[width=\linewidth]{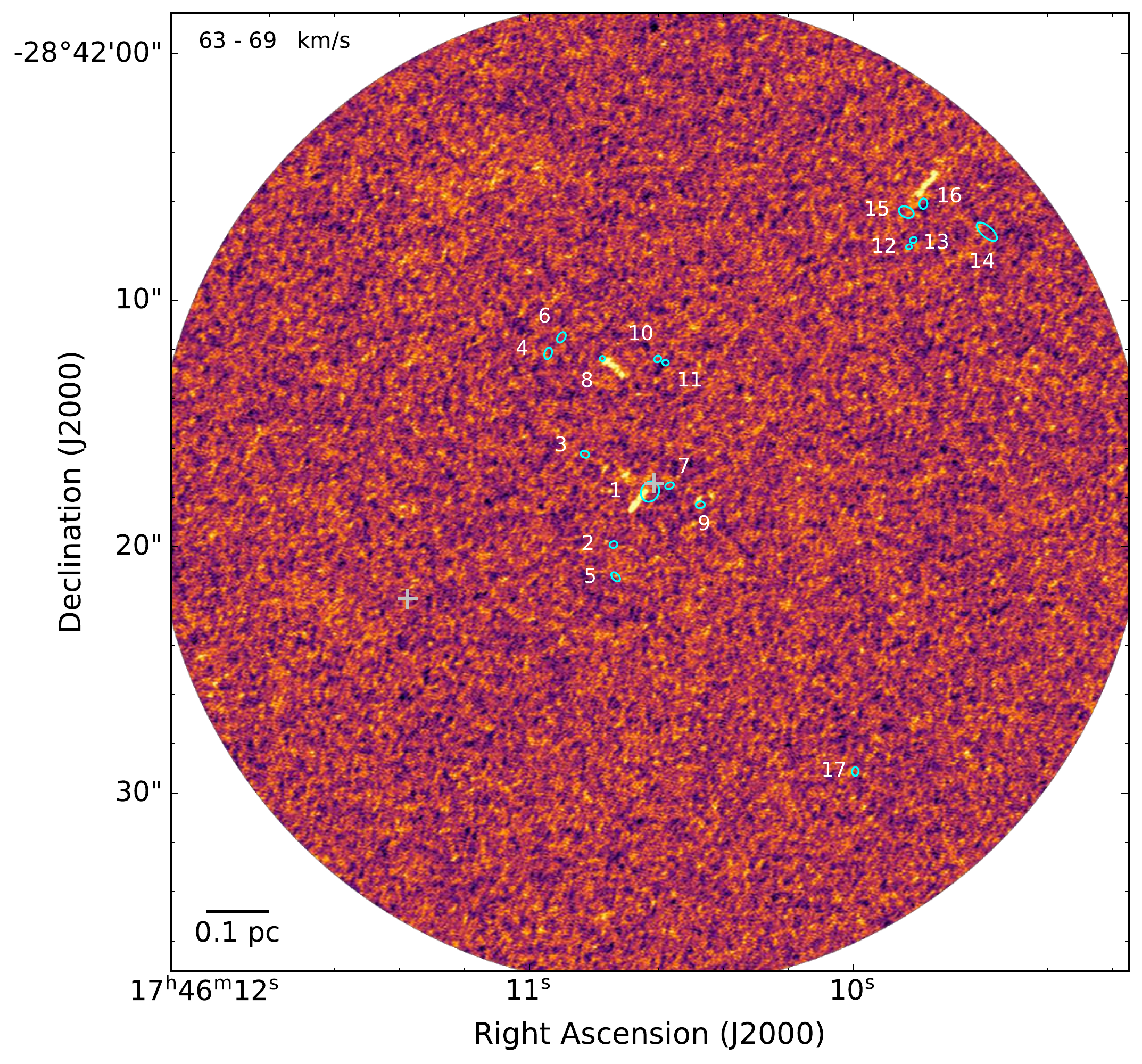}
    \caption{} 
\end{subfigure}%
\hfill
\begin{subfigure}[b]{0.49\textwidth}
    \centering
    \includegraphics[width=\linewidth]{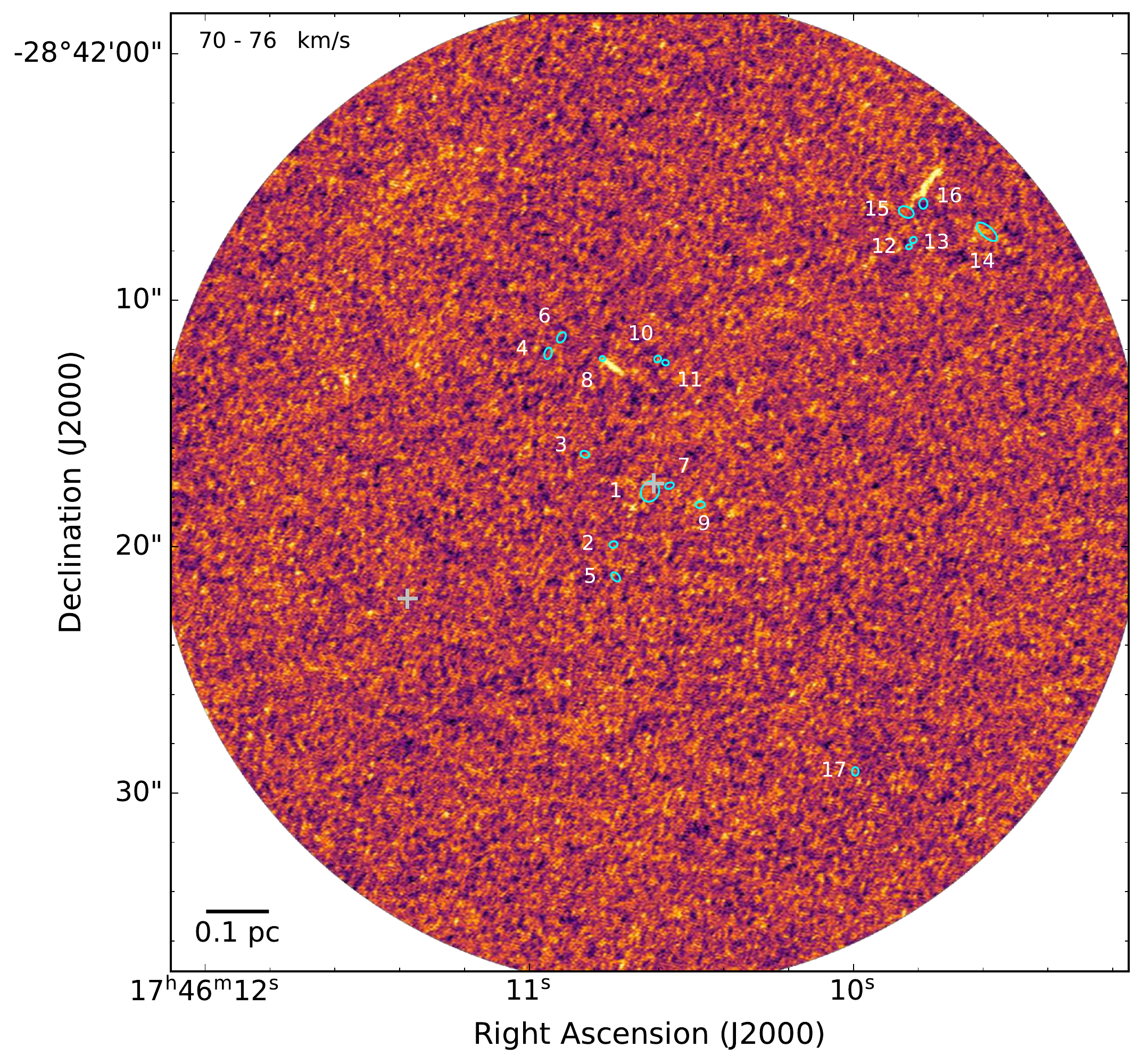}
    \caption{} 
    \end{subfigure}
    
\begin{subfigure}[c]{0.49\textwidth} 
    \includegraphics[width=\linewidth]{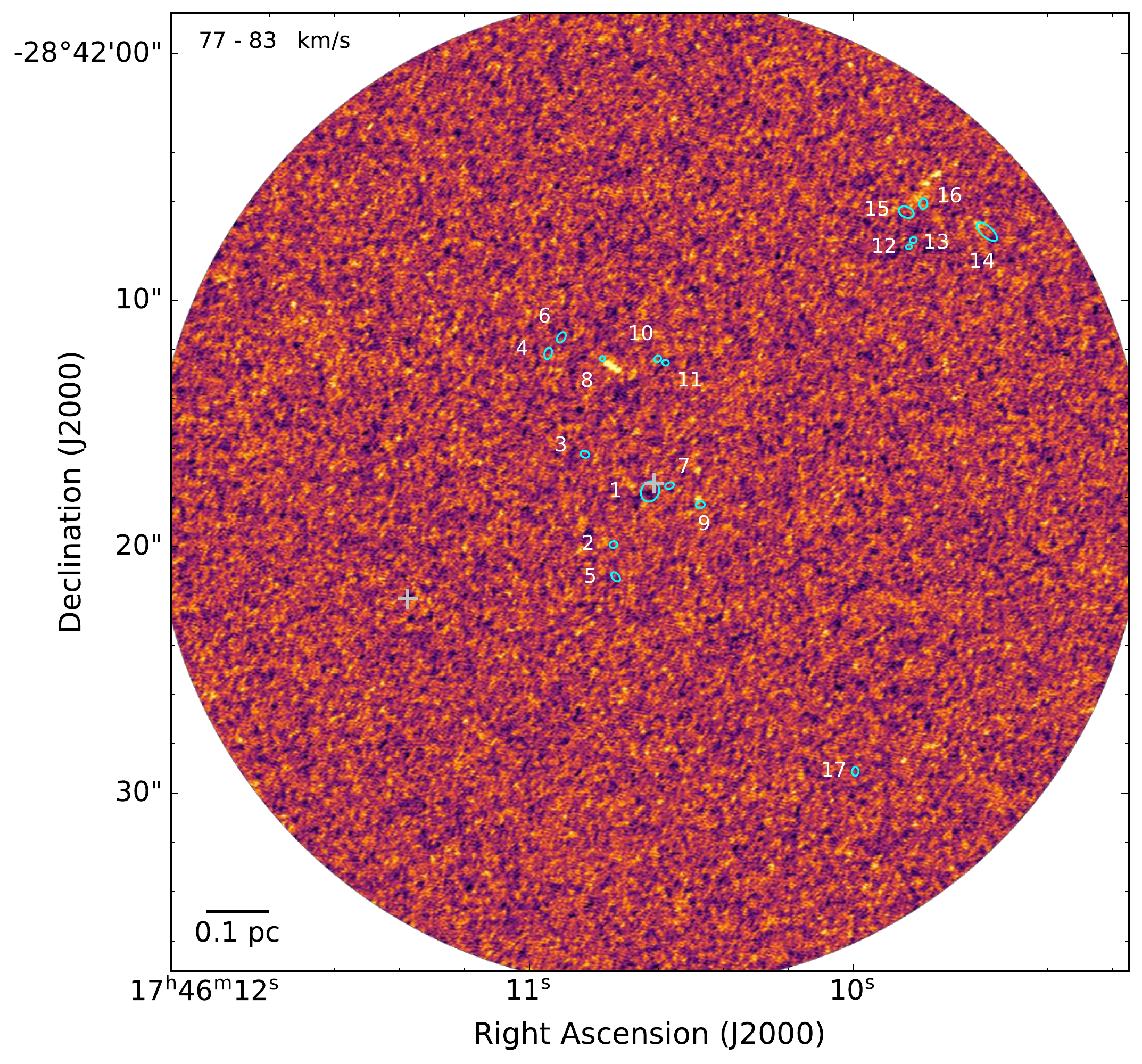}
    \caption{} 
\end{subfigure}%
    \caption{As in Figure 10. \textbf{(a)}: 63 -- 69 \kms, \textbf{(b)}: 70 -- 76 \kms, \textbf{(c)}: 77 -- 83 \kms.} 
\label{fig:SiO_appendix5}
\end{figure*}

\clearpage
\onecolumn
\noindent
\section*{Author affiliations}
$^{1}$Department of Physics, University of Connecticut, 196A Auditorium Road, Storrs, CT 06269 USA\\
$^{2}$National Astronomical Observatory of Japan, 2-21-1 Osawa, Mitaka, Tokyo, 181-8588, Japan\\
$^{3}$Joint ALMA Observatory, Alonso de C\'{o}rdova 3107, Vitacura, Santiago, Chile\\
$^{4}$Astrophysics Research Institute, Liverpool John Moores University, IC2, 146 Brownlow Hill, Liverpool, L3 5RF, United Kingdom\\
$^{5}$CASA, University of Colorado, 389-UCB, Boulder, CO 80309\\
$^{6}$Department of Astronomy, University of Florida, PO Box 112055, Gainesville, FL 32611, USA\\
$^{7}$Astronomisches Rechen-Institut, Zentrum f\"{u}r Astronomie der Universit\"{a}t Heidelberg, M\"{o}nchhofstra\ss e 12-14, 69120 Heidelberg, Germany\\
$^{8}$Center for Astrophysics $\vert$ Harvard \& Smithsonian, 60 Garden Street, Cambridge, MA 02138, USA\\
$^{9}$Max-Planck Institute for Astronomy, K\"{o}nigstuhl 17, 69117 Heidelberg, Germany\\
$^{10}$University of Vienna, Department of Astrophysics, T\"{u}rkenschanzstrasse 17, 1180 Wien, Austria\\ 
$^{11}$Radcliffe Institute for Advanced Study, Harvard University, 10 Garden Street, Cambridge, MA 02138, USA\\
$^{12}$Argelander-Institut f\"{u}r Astronomie, Universit\"{a}t Bonn, Auf dem H\"{u}gel 71, 53121, Bonn, Germany\\
$^{13}$Leiden Observatory, Leiden University, PO Box 9513, NL 2300 RA Leiden, the Netherlands\\
$^{14}$Kavli Institute for Astronomy and Astrophysics, Peking University, Beijing 100871, China\\
$^{15}$Department of Astronomy, School of Physics, Peking University, Beijing 100871, China\\
$^{16}$SOFIA Science Center, USRA, NASA Ames Research Center, Moffett Field CA 94045, USA\\
$^{17}$Haystack Observatory, Massachusetts Institute of Technology, 99 Millstone Road, Westford, MA 01886, USA\\
$^{18}$Department of Physics and Astronomy, University of Kansas, 1251 Wescoe Hall Drive, Lawrence, KS 66045, USA\\
$^{19}$Boston University Astronomy Department, 725 Commonwealth Avenue, Boston, MA 02215, USA\\

\bsp
\label{lastpage}

\end{document}